\documentclass[11pt]{article}

\usepackage{amssymb}
\usepackage{amsmath}
\usepackage{amsthm}
\usepackage{multirow}
\usepackage{color, xcolor}
\usepackage{fullpage}
\usepackage{makeidx}
\makeindex

\usepackage[utf8]{inputenc}
\usepackage[margin=3cm]{geometry}
\usepackage{natbib}
\usepackage{url}
\usepackage{bm}
\usepackage{array}
\usepackage{hyperref}

\usepackage{setspace}
\usepackage{booktabs}
\usepackage{tabularx}

\usepackage{graphicx}
\usepackage{caption}
\usepackage{subcaption}

\definecolor{bluegreen}{RGB}{0, 100, 150}
\definecolor{red3}{RGB}{250, 20, 0}
\definecolor{green4}{RGB}{58, 198, 91}
\definecolor{yellow5}{RGB}{180, 180, 0}
\definecolor{orange6}{RGB}{255, 150, 0}
\definecolor{blue7}{RGB}{10, 50, 250}
\definecolor{purple8}{RGB}{130, 110, 190}
\definecolor{green9}{RGB}{0, 90, 0}

\graphicspath{{./Figures/}}

\DeclareMathOperator*{\argmin}{arg\,min}
\RequirePackage[normalem]{ulem} 
\RequirePackage{color}\definecolor{RED}{rgb}{1,0,0}\definecolor{BLUE}{rgb}{0,0,1} 
\providecommand{\DIFaddbegin}{} 
\providecommand{\DIFaddend}{} 

\providecommand{\keywords}[1]
{
	\small	
	\textbf{\textit{Keywords---}} #1
}

\begin{document}

	\title{Estimating Individual Treatment Effects using Non-Parametric Regression Models: a Review}

\author{%
	Alberto Caron \thanks{
		This work was supported by a British Heart Foundation-Turing Cardiovascular Data Science Award (BCDSA/100003).
		Corresponding author: \texttt{alberto.caron.19@ucl.ac.uk}, 1-19 Torrington Pl, London WC1E 7HB. } \\
	\footnotesize Department of Statistical Science\\
	\footnotesize University College London\\
	\and
	Gianluca Baio \\
	\footnotesize Department of Statistical Science\\
	\footnotesize University College London\\
	\and
	Ioanna Manolopoulou \\
	\footnotesize Department of Statistical Science\\
	\footnotesize University College London\\
}

	\maketitle
	\setcounter{page}{1}
	
	\begin{abstract}
		
		Large observational data are increasingly available in disciplines such as health, economic and social sciences, where researchers are interested in causal questions rather than prediction.
		In this paper, we examine the problem of estimating heterogeneous treatment effects using non-parametric regression-based methods, starting from an empirical study aimed at investigating the effect of participation in school meal programs on health indicators. Firstly, we introduce the setup and the issues related to conducting causal inference with observational or non-fully randomized data, and how these issues can be tackled with the help of statistical learning tools. Then, we review and develop a unifying taxonomy of the existing state-of-the-art frameworks that allow for individual treatment effects estimation via non-parametric regression models. After presenting a brief overview on the problem of model selection, we illustrate the performance of some of the methods on three different simulated studies. We conclude by demonstrating the use of some of the methods on an empirical analysis of the school meal program data.

	\end{abstract}
	

	
	\section{Introduction}  \label{sec:intro}
	The application of advanced statistical learning tools in causal inference has gained popularity in recent years, partly due to the fact that large datasets are becoming available at relatively lower costs (thanks to e.g.~electronic health records, social network data etc.). One of the increasingly common objectives of causal inference in many disciplines is to draw inferences about individual-level treatment effects, as opposed to inferring treatment effects on average across the entire population. The importance of inferring individual-level treatment effects lies in the fact that treatment effects are very often heterogeneous across units of analysis. Two such examples arise in precision medicine and personalized advertisement, where the ultimate goal is to make decisions at the level of an individual patient or user \citep{persmark, precmed}. For instance, patients with high cholesterol levels respond differently to statin prescriptions based on their clinical records. This level of analysis requires causal inference methods that can accurately predict the impact of treatment, as well as quantify its uncertainty, at a fine resolution. To this end, popular statistical learning algorithms such as tree ensembles \citep{athey_2019, hahn_2020}, kernel methods \citep{vanderschaar_2017, vanderschaar_2018} or neural networks \citep{johansson_2016, shalit_2017}, that exhibit excellent performance in capturing complex non-linear relationships, can be exploited also in causal settings after due adjustments. This paper is therefore motivated by the growing number of non-parametric regression-adjustment methods for modelling \emph{Individual Treatment Effects} (ITE) using large datasets to answer individual-level causal questions, and reviews some of the most popular meta-algorithm frameworks that allow to do that, laying down the underlying assumptions and pros/cons for each of them.

	In order to better explain the rationale and the challenges behind estimation of heterogeneous treatment effects, we refer to an applied study in the social sciences, first analyzed by \cite{chan_2016}. The dataset consists of a subset of the 2007-2008 National Health and Nutrition Examination Survey (NHANES)\footnote{\url{https://www.cdc.gov/nchs/nhanes/index.htm}.} of children aged 4-17, aimed at investigating whether participation in the National School Lunch or the School Breakfast programs is associated with improved children's health in low-income households, measured through body-mass index. The ex-post study of treatment effect heterogeneity in this case is important to identify children who benefited the most and the least from the program, and to investigate whether there is evidence of negative or null treatment impact for some specific subgroups of children. A further challenge for policy design and improvement is then related to identifying the main factors driving heterogeneity behind treatment effects (i.e.~moderators). For instance, one might suspect that age is a primary treatment effect moderator and would like to provide evidence in favour or against this hypothesis. Understanding heterogeneity of treatment is crucial in guiding future policy interventions. In later sections, we will frame the problem of studying heterogeneity and moderation in treatment effects more formally, and finally demonstrate the use of some empirical methods on the NHANES data to answer these type of questions.

	Regardless of the precise causal questions of interest, the fundamental challenge in causal inference is that the quantity of interest  --- namely the effect of treatment --- depends on an unobservable counterfactual. Moreover, in non-randomized studies, treatment effect and selection into treatment are invariably entangled. As a result, advanced statistical learning tools, capable of exploiting large datasets in supervised learning settings, can be directly applied, but easily stumble into bias issues. Instead, the latent treatment effect is inferred by reconstructing counterfactual statements either through sampling (randomization in the administration of treatment) or by adjusting for covariates that affect both the outcome of interest and treatment assignment, and thus ``confound'' the treatment effect (confounders). In the NHANES data example, confounding might be associated for instance with children's age or ethnic background, both representing potential common causes of body-mass index and participation in the school meal program.
	
	Randomized experiments, where treatment is randomly allocated, marginally on confounding factors (such as medical or socio-economic characteristics), are considered to be the gold standard for causal inference, as they are designed to control for confounding and to offer a good approximation of counterfactuals (e.g.~placebo controlled trials). However, fully randomized studies are often costly, difficult to access, and sometimes suffer from problems such as non-compliance and other issues of missingness not at random that might invalidate the randomization mechanism, and external validity of the results.
	
	In contrast, data of observational nature, where treatment administration is not randomized, are more easily accessible and abundantly present in many applied fields. However, observational studies present several drawbacks, largely attributable to three complex phenomena. The first is \textit{selection bias}, which manifests when the treatment allocation mechanism is not under the researcher's control, but determined by observable or unobservable factors. This constitutes a potential source of confounding that needs to be controlled for, as it generates structural differences between the treated and the control groups. The second phenomenon is known as \textit{partial overlap}, which occurs when there are regions in the space of relevant covariates where only treated, or only control, units are present. As a result, units in these particular regions lack an appropriate comparator with similar covariates. These two issues are closely related, as partial overlap may be a direct consequence of selection bias. The third phenomenon relates to the fact that treatment allocations and their corresponding outcomes may not be independent across individuals.

	In this paper we review some of the most recent frameworks that allow to use non-parametric regression models for estimating heterogeneous effects arising from a binary treatment assignment (\cite{vanderschaar_2017, athey_2019, wager_2019, hahn_2020} among others), at a individual level, and introduce a taxonomy that classifies these methods under the same unified framework. More in details, we provide an overview of the implied assumptions and compare methods' performance with the help of two simulated studies. In addition, we illustrate a practical application of some of the methods by analyzing the NHANES data, with the aim of investigating the presence of heterogeneous effects of school meal participation on children's BMI, and detecting the moderators responsible for heterogeneity. The ultimate goal of this paper is to expand previous review works, such as \cite{kunzel_2017} and \cite{knaus_2020}, by constructing a high-level taxonomy of regression-based methods for ITE estimation and 
		including some of the most recent additions to the literature, including those coming from computer science fields, e.g.~\cite{johansson_2016, shalit_2017, vanderschaar_2017, vanderschaar_2018}.
	
	We focus on methods which circumvent the potential issues stemming from lack of complete (and controlled) randomization by making a set of assumptions which are common in the literature. The first assumption is \emph{unconfoundedness}, which ensures that there is no unobserved confounding factor driving selection into treatment. Unconfoudedness is untestable, and might be in some cases a strained assumption to make, but represents less of a threat in settings where a large number of relevant covariates is available, deemed to be good proxies of confounders. The second assumption is \emph{common support}, which states that each unit, identified by a given set of covariates, has non-zero probability of being observed in each of the treatment groups. In other words, common support ensures that there is no deterministic component in treatment administration. Contrary to unconfoundedness, common support can be inspected in the data. Finally, the last assumption is known as \emph{Stable Unit Treatment Value Assumption} (SUTVA), and states that the response to treatment of one unit is not affected by other units' assignment to treatment, thus ensuring that there is no interference. 
	
	The rest of the work is organized as follows. Section \ref{sec:ProblemSetup} formulates the problem of estimating \emph{Conditional Average Treatment Effect} (CATE) by using the potential outcomes representation by \cite{rubin1978}. Section \ref{sec:cate} reviews the most popular frameworks for estimating CATE and comments on the problem of model selection. Section \ref{sec:simu} presents results from two different simulated studies that we conducted to compare performance of some of the methods. Section \ref{sec:realdata} provides an example of a real-world social sciences application. Section \ref{sec:discussion} concludes with a discussion.

	\section{Problem setup}\DIFaddbegin \label{sec:ProblemSetup}

We will follow the Neyman-Rubin Causal Model, outlined in \cite{rubin1978} and \cite{imbens_rubin_2015}, which conceives causal inference as a missing data problem. For each unit of analysis $i \in \{ 1, ..., N \}$, given a binary treatment assignment $Z_i \in \{ 0,1 \}$, where $Z_i = 1$ indicates exposure to the treatment and $Z_i = 0$ indicates no exposure, the framework defines the quantities $\big( Y_i^{(0)}, Y_i^{(1)} \big)$ as potential outcomes. $Y_i^{(1)}$ corresponds to unit $i$'s outcome under exposure to the treatment, thus in our NHANES study to the child's BMI given participation in school meals; while $Y_i^{(0)}$ corresponds to its outcome under no exposure, namely BMI in case of no participation in school meals. Only one of these two is actually observed. We will consider, throughout this work, a setting where the outcome variable of interest is continuous, i.e.,~$\big( Y_i^{(0)}, Y_i^{(1)} \big) \in \mathbb{R}^2$. However most of the ideas and methods presented can be generalized to multiple treatment arms and binary or count outcomes. 

It is worth pointing out that the Neyman-Rubin Causal Model (RCM) is not the only viable framework for causal inference. For example, \cite{pearl_2009} offers a more graphical approach based on Causal Directed Acyclic Graphs (DAGs) \citep{tennant_2020} and \emph{do}-calculus to represent counterfactuals; while \cite{dawid_2000, dawid_2015} instead develops a critique of counterfactuals and proposes a different framework that does not rely on them, based on Bayesian decision analysis. We stress that the same structural causal model, made of a set of (non-parametric) structural equations, admits equal representation under all the different causality frameworks mentioned above. Thus, albeit we explicitly employ RCM to represent the problem of ITE estimation, the methods and estimators presented in Section \ref{metalearners} are general and valid under other causal models.

	Given the framework outlined above, in case both potential outcomes were observed, one would have access also to the \emph{Individual Treatment Effects} (ITE), defined as $Y_i^{(1)} - Y_i^{(0)}$. However, as described earlier, the fundamental problem of causal inference is that, for each individual $i$, only one of the two potential outcomes $\big( Y_i^{(0)}, Y_i^{(1)} \big) \in \mathbb{R}^2$ is observed, corresponding to the realized treatment assignment: $Y_i = Z_i Y_i^{(1)} + (1 - Z_i) Y_i^{(0)}$. This implies that $Y_i^{(1)} - Y_i^{(0)}$ cannot be estimated as in a usual regression problem.
	
	Given a dataset of either observational or randomized nature $\mathcal{D}_i = \{ \bm{X}_i, Z_i, Y_i \} $, with $i \in \{ 1, ..., N \}$, where $\bm{X}_i \in \mathcal{X}$ denotes a $d$-dimensional set of observed covariates for the individual $i$ which are potential source of confounding to be controlled for, the aim is to estimate Conditional Average Treatment Effect (CATE), defined as
	\begin{equation} \label{CATE}
	\tau(\bm{x}_i) = \mathbb{E} \Big[ Y_i^{(1)} - Y_i^{(0)} \mid \bm{X}_i = \bm{x}_i \Big]~.
	\end{equation}

	\begin{figure}[t]
		\centering
		\begin{minipage}{.5\textwidth}
			\centering
			\includegraphics[scale=0.52]{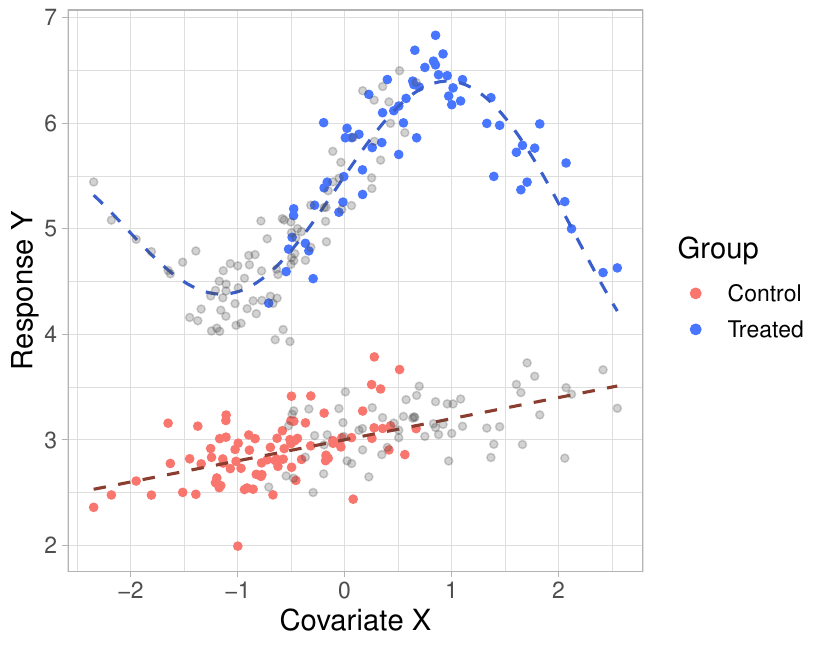}
		\end{minipage}%
		\begin{minipage}{.4\textwidth}
			\centering
			\includegraphics[scale=0.52]{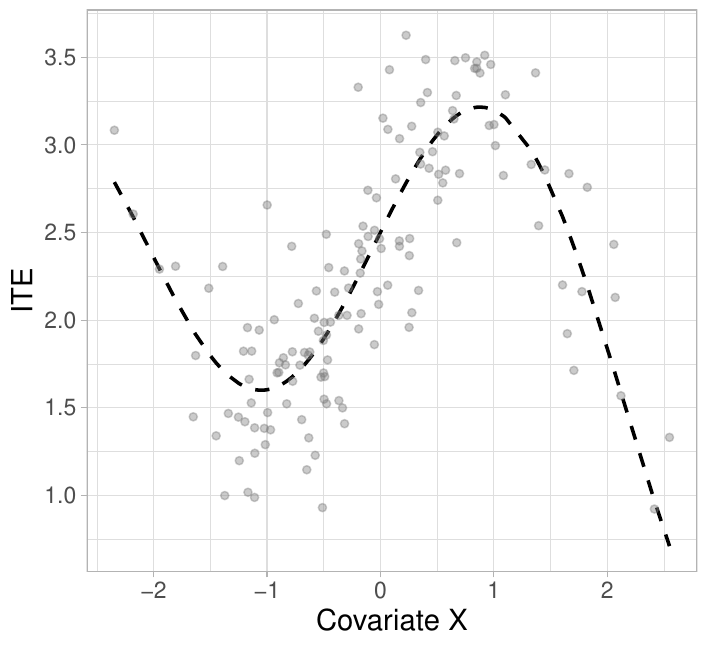}
		\end{minipage}
		\caption{\small Simulated example with one single covariate $X$. Potential outcomes are generated respectively as $Y_i^{(0)} \sim \mathcal{N} (3 + 0.2 X_i, 0.25)$ and $Y_i^{(1)} \sim \mathcal{N} (5.5 - 0.1X_i^2 + \sin (1.5X_i), 0.25)$, while the propensity score is $\pi(X_i) = \Phi (-0.2 + 1.5 X_i)$, where $\Phi(\cdot)$ is the standard normal cdf. Left panel: observed outcomes for the treated (blue dots) and control (red dots) groups with underlying conditional mean function (dashed lines), and unobservable counterfactual outcomes (grey dots). Right panel: unobservable true ITE (grey dots) and corresponding conditional mean function (dashed line).}
		\label{example}
	\end{figure}

	\noindent The two quantities $\mu_Z (\bm{x}_i) = \mathbb{E} \big[ Y^{(Z_i)} \mid \bm{X}_i = \bm{x}_i \big]$ in (\ref{CATE}) are the conditional average potential outcomes. The intuition behind the estimation of $\tau(\bm{x}_i)$ is the following. In case both potential outcomes were observed, then $Y_i^{(1)} - Y_i^{(0)}$ (ITE) would be modelled as the response variable in a regression framework where $\bm{X}_i$ are the $d$ regressors, and where the aim is to estimate the conditional mean of the outcome, namely $\tau(\bm{x}_i) = \mathbb{E} [ Y_i^{(1)} - Y_i^{(0)} \mid \bm{X}_i = \bm{x}_i ]$. The set of regressors $\bm{X}_i$ here does not necessarily include all the available covariates, but only moderators of treatment effects responsible for heterogeneity in the response. We will discuss in later sections how detecting moderators is a particularly insightful part of the analysis of heterogeneous treatment effects. Figure \ref{example} provides a graphical representation of a simple single-covariate example, where coloured dots show observed values $Y_i = Y^{(Z_i)}$ of the response, while grey dots their corresponding (unobservable) counterfactuals $Y_i^{(1-Z_i)}$. Notice that the example is purely illustrative and serves as visual aid to introduce the reader to the key concepts in the Rubin-Neyman framework.
	
	An additional quantity of interest under the Neyman-Rubin framework is the \emph{propensity score}, which is defined, for each unit of analysis $i$, as the probability of being selected into treatment, given a set of observed covariates. As also mentioned earlier with regards to ITE moderators, the set of covariates determining the propensity score is not necessarily the same as that determining $Y_i$, or the same as the moderators' set. Nevertheless, the narrowest set of covariates needed to achieve unbiased treatment effect estimates is represented by the confounders, which have to be included in both propensity and outcome models. The propensity score is formally defined as:
	\begin{equation*}
	\pi(\bm{x}_i) = \mathbb{P} ( Z_i = 1 \mid \bm{X}_i = \bm{x}_i ) ~ .
	\end{equation*}
	Thus, for each unit, the binary treatment assignment $Z_i$ can be seen as the outcome of a Bernoulli experiment where $Z_i \sim \text{Bern} \big( \pi(\bm{x}_i) \big)$. In the NHANES observational study, the propensity score is unknown by construction, and if we are willing to employ CATE estimation methods that are based on the propensity score, we have to estimate it. In constructing a propensity score model, we expect age and ethnic background to be the main determinants of selection into treatment. In the simple example of Figure \ref{example}, the propensity score is generated as a monotone function of the covariate $X$; this is why treated units are more frequently observed for higher values of $X$, while control units are more frequent for lower values of $X$. The propensity score distribution in this case is also slightly skewed to the right; as a consequence, there are more units in the control group compared to the treated one. 
	
	Under the potential outcomes framework, the untestable unconfoundedness assumption is equivalent to assuming the conditional independence 
	\begin{equation} \label{CIA}
	\big( Y_i^{(0)}, Y_i^{(1)} \big) \perp \!\!\! \perp Z_i \mid \bm{X}_i  ~ ,
	\end{equation}
	which ensures that all the common causes of $Y_i$ and $Z_i$ are observed. A well-known result concerning the propensity score, derived by \cite{rosenbaum_1983}, is that if (\ref{CIA}) holds, then
	\begin{equation} \label{CIAps}
	\big(Y_i^{(0)}, Y_i^{(1)} \big) \perp \!\!\! \perp Z_i \mid \pi(\bm{X}_i)~.
	\end{equation}
	Conditional independence in (\ref{CIAps}) represents a different way of expressing unconfoundedness, by conditioning only on the (true) propensity score rather than on the full set of covariates. In the context where the number of covariates is high, $\pi(\bm{X}_i)$ represents a 1-dimensional summary of a $d$-dimensional covariate set $\bm{X}_i \in \mathcal{X}$ which achieves conditional independence between outcome and treatment assignment. There are many different ways of making effective use of the propensity score for deriving population-level or individual-level causal estimators. In some approaches the propensity score assumes a central role, such as in matching and blocking methods \citep{rosenbaum_1984, rosenbaum_1985}, compared to regression-based methodologies that we focus on in this work. \cite{imbens2004} offers a nice and thorough overview of methods based on the propensity score, and also of other methods (e.g.~regression and matching). In observational studies, where the propensity score is unknown and to be estimated, the success of propensity methods heavily depends on the quality of the specified propensity model. For instance, a poor approximation of the propensity scores might induce instability and high variance when it is employed for re-weighting observations (e.g.~presence of near zero propensities in the Inverse Propensity Weighting scheme).
	
	The assumption of common support (or overlap), under the Rubin-Neyman framework, implies that the propensity score is such that $0 < \pi(\bm{x}_i) < 1$, for each $i$, which equivalently means that treatment assignment is not deterministic, and that an individual with covariate values $\bm{X}_i = \bm{x}_i$ can be potentially observed in either treatment group with non-zero probability. Notice that common support does not necessarily need to hold for all the available covariates in the covariate set $\bm{X}_i$, but just for the confounders, which might constitute a strict subset of $\bm{X}_i$. For this reason common support is sometimes referred to as \emph{common causal support} \citep{hill_2013}. Common support may hold only for a portion of the available sample as we discuss in more details in the last paragraph of this section, so that inference on treatment effects outside the guaranteed overlap region becomes unreliable. Fortunately, contrary to unconfoundedness, there are effective ways to empirically check whether common support holds in the data. However, inspection of common support regions with more naive methods such as visual inspection might be challenging when $\bm{X}_i$ is high-dimensional. Some Bayesian non-parametric implementations of CATE estimation models offer a simple yet effective way of checking for common support regions, as described in \cite{hill_2013}. In the simple one-covariate example of Figure \ref{example}, the propensity score takes values which are very close to either 0 or 1, but it is guaranteed to lie strictly between the two in the data generating process. This implies that overlap theoretically holds on the whole $\mathcal{X}$ support, but due to sample finiteness we are not able to capture enough variation and we observe only treated units for higher values of $X_i$ and only controls for lower values of $X_i$; we would consequently proceed by narrowing down the analysis to the overlap region, i.e.~in a neighborhood of $X_i = 0$.
	
	Under unconfoundedness and common support assumptions, CATE can be estimated from observed data in $\mathcal{D}_i = \{ \bm{X}_i, Z_i, Y_i \}\;i \in \{ 1, ..., N \} $. In fact, under unconfoudedness, the conditional average potential outcomes are such that
	\begin{equation}  \label{mu_ide}
	\begin{split}
	\mu_{Z} (\bm{x}_i) & = \mathbb{E} \left[ Y^{(Z_i)} \mid \bm{X}_i = \bm{x}_i \right] = \mathbb{E} \left[ Y^{(Z_i)} \mid Z_i = z_i, \bm{X}_i = \bm{x}_i \right] \\
	& = \mathbb{E} \left[ Y_i \mid Z_i = z_i, \bm{X}_i = \bm{x}_i \right]~,
	\end{split}
	\end{equation}
	for $z_i \in \{ 0, 1 \}$, where the second equality generates from the conditional independence between $\big( Y_i^{(0)}, Y_i^{(1)} \big)$ and $Z_i$ given $\bm{X}_i$, while the last one from the identity $Y_i = Z_i Y_i^{(1)} + (1 - Z_i) Y_i^{(0)}$. Hence, as a straightforward implication of (\ref{mu_ide}), one can derive an estimator for CATE as
	\begin{equation} \label{CATEide}
	\begin{split}
	\tau(\bm{x}_i) & = \mathbb{E} \left[ Y_i^{(1)} - Y_i^{(0)} \mid \bm{X}_i = \bm{x}_i \right]
	\\
	& = \mathbb{E} \left[ Y_i^{(1)} \mid \bm{X}_i = \bm{x}_i \right] - \mathbb{E} \left[ Y_i^{(0)} \mid \bm{X}_i = \bm{x}_i \right]
	\\
	& = \mathbb{E} \left[ Y_i^{(1)}  \mid Z_i = 1 , \bm{X}_i = \bm{x}_i \right] - \mathbb{E} \left[ Y_i^{(0)} \mid Z_i = 0 , \bm{X}_i = \bm{x}_i \right]
	\\
	& = \mathbb{E} \left[ Y_i \mid Z_i = 1 , \bm{X}_i = \bm{x}_i \right] - \mathbb{E} \left[ Y_i \mid Z_i = 0 , \bm{X}_i = \bm{x}_i \right]~,
	\end{split}
	\end{equation}
	where, as in (\ref{mu_ide}), the third equality is given by unconfoundedness, and the last one by the identity $Y_i = Z_i Y_i^{(1)} + (1 - Z_i) Y_i^{(0)}$. Common support assumption is then needed to guarantee that the two conditional average potential outcomes $\mu_{Z} (\bm{x}_i) = \mathbb{E} \left[ Y^{(Z_i)} \mid \bm{X}_i = \bm{x}_i \right]$ exist for each values $z_i$ and $\bm{x}_i$ in their supports, and thus can be estimated through the observed quantities in the conditional expectations $\mathbb{E} \left[ Y_i \mid Z_i = z_i, \bm{X}_i = \bm{x}_i \right]$. To clarify, suppose that common support does not hold for $\bm{X}_i = x^*$ and that $\pi (x^*) = 0$ (without loss of generality), then the conditional average potential outcome $\mu_1 (x^*) = \mathbb{E} \big[ Y^{(1)} \mid \bm{X}_i = x^* \big]$ does not theoretically exist, and it would not make sense to attempt to even estimate it.
	
	In empirical studies of observational nature, common support sometimes fails to hold for the whole support of $\bm{X}_i$. Hence, CATE can only be identified and reliably estimated in some specific sample regions. In particular, we might refer to estimands such as \emph{Average}, or \emph{Conditional Average}, \emph{Treatment Effect on the Treated} (ATT and CATT, respectively) to indicate when treatment effects are identifiable only on the treated group, or equivalently when common support only holds for the treated units. The same holds for \emph{Average}, or \emph{Conditional Average}, \emph{Treatment Effect on the Control} (ATC and CATC). In practice, empirical methods such as propensity score re-weighting and trimming are generally utilized in order to focus estimation on overlap regions only.
	
	\subsection{Non-Parametric regression framework for CATE estimation}	
	
	Throughout this work, we will review non-parametric regression approaches where the outcome surface $Y_i$ is modelled as a function of $( \bm{X}_i$, $Z_i )$ and some unobservable error term $\varepsilon_i$~: $Y_i = g (\bm{X}_i, Z_i, \varepsilon_i)$. More specifically, the reviewed methods generally assume that the error term $\varepsilon_i$ is additive and with mean zero, which leads to the following setup:
	\begin{equation} \label{onestage}
	Y_i = f(\bm{X}_i, Z_i) + \varepsilon_i~, \qquad \text{where} \quad  \mathbb{E}(\varepsilon_i) = 0~,
	\end{equation}
	and $f(\bm{X}_i, Z_i) = \mathbb{E} \big[ Y_i \mid \bm{X}_i, Z_i \big]$ is left unspecified and learnt from the data. The strength of non-parametric regression models is that they are less prone to misspecification of the functional form of $f(\cdot)$ (e.g.~tree-based methods model $f(\cdot)$ as piece-wise constant, splines as piece-wise polynomial, etc.). As mentioned earlier in the section, the covariates $\bm{X}_i \in \mathcal{X}$ represent a potential source of confounding to be controlled for. In model building terms, this means that ideally confounding variables contained in $\bm{X}_i$ need to be included in both propensity and outcome models, while different subsets of other covariates might be included in the propensity and outcome models if they're relevant predictors of either $Z_i$ or $Y_i$, albeit not common causes. In addition, in a setting where the number of available covariates is high, one might want to resort to regularization in the estimation of $f(\cdot)$. However, as explained in both \cite{hahn_2018} and \cite{hahn_2020}, regularization should be handled carefully in this context; we will return to this point in Section \ref{sec:taulearner}.
	
	In the regression setup illustrated in (\ref{onestage}), some of the frameworks reviewed in the next section reserve a specific role for the propensity score (X-, R- and $\tau$-Learners) --- these are often referred to as ``propensity methods". For the remaining methods which do not explicitly envisage the use of the propensity score, in the simulated studies that we conduct later in Section \ref{sec:simu}, we follow the suggestion of \cite{hahn_2020} and incorporate PS estimates as an additional covariate, according to the following two-stage regression framework:
	\begin{align} \label{twostage}
	\begin{split}
	\pi(\bm{X}_i) = & ~ \mathbb{P} \big( Z_i = 1 \mid \bm{X}_i \big)   \\
	Y_i = & ~ f \Big( \big[ \bm{X}_i ~~ \pi(\bm{X}_i) \big] , Z_i \Big) + \varepsilon_i ~ .
	\end{split}
	\end{align}
	The first stage in (\ref{twostage}) involves estimating the propensity score, while the second embeds it as an extra covariate in the covariate set. Any probabilistic classifier is suitable for use in the first stage regression (e.g.~logistic regression, deep neural networks, etc.). As explained in \cite{hahn_2020}, and as we will describe later in Section \ref{sec:taulearner}, the addition of $\pi(\bm{X}_i)$ to the covariate set in (\ref{twostage}) represents an effective way to tackle bias deriving from \emph{targeted selection}. Targeted selection arises when individuals are selected into treatment based on the prediction of otherwise adverse outcome (or of large gains under treatment), i.e.,~when $\pi(\bm{X}_i)$ is a strictly monotone function of $\mathbb{E} [ Y_i^{(0)} \mid \bm{X}_i = \bm{x}_i ]$, and is common in many observational studies (e.g.~medical or socio-economic studies). The reason why we decided to add PS estimates as an extra covariate in those non-propensity methods is that we aim to provide a comparison of methods' performance in simulation studies (Section \ref{sec:simu}) which is exclusively based on the way they derive a CATE estimator, and decouple any difference in performance attributable to the propensity score inclusion in the model.
	
	CATE estimators can be directly derived from the representations in (\ref{onestage}) and (\ref{twostage}). There are currently few different approaches for deriving an estimator for CATE from (\ref{onestage}) and (\ref{twostage}), that will be analyzed in the next section.

	\section{Estimating CATE} \label{sec:cate}

    Given the framework outlined in the previous section, various meta-algorithms designed to derive a CATE estimator have been proposed in the literature. These meta-algorithms are often referred to as ``Meta-Learners'', in that they are subroutines of ``base-learners'', which are common machine learning algorithms (e.g.~tree ensembles, neural networks, gradient boosting methods, etc.). We attempt to build a unifying taxonomy of these ``Meta-Learners'' approaches in Section \ref{sec:metalearners}, while in Section \ref{sec:modelsel} we present an overview on the problem of model selection for CATE estimation, which is a substantially hard, arguably impossible, problem.

	\subsection{Meta-Learners} \label{sec:metalearners}
	
	As mentioned in the earlier sections, we will partly build on top of the work by \cite{kunzel_2017} and expand it by including the most recent contributions stemming from both the statistics and computer science literature. A concise summary of the presented ``Meta-Learners'', together with the relevant references, can be found in Table \ref{metalearners}.
	
	\begin{table}
		\caption{Summary of meta-learners described in this paper}
		\footnotesize
		\centering
		\begin{tabular}{>{\small\arraybackslash}m{3.3cm} | >{\small\arraybackslash}m{4cm} | >{\small\arraybackslash}m{6.1cm}}
			&   \textbf{References}   &    \textbf{CATE estimator}  
			\\
			\midrule
			
			\textbf{S-Learner} &  \cite{hill_2011, foster_2011}  &   $\tau (\bm{x}_i) = f(\bm{x}_i, 1) - f(\bm{x}_i, 0)$
			\\
			\midrule
			
			\textbf{T-Learner} &  \cite{athey_2016, lu_sadiq_2018}, \cite{powers_2018}  &   $\tau (\bm{x}_i) = f_1(\bm{x}_i) - f_0(\bm{x}_i)$
			\\
			\midrule

			\textbf{X-Learner} &  \cite{kunzel_2017}  &  $ \tau(\bm{x}_i) = \pi(\bm{x}_i) \tau_0(\bm{x}_i) + \big(1 - \pi(\bm{x}_i) \big) \tau_1(\bm{x}_i) $
			\\
			\midrule
			
			\textbf{R-Learner} &  \cite{wager_2019}  &  $  \tau(\bm{x}_i) = \argmin_{\tau} \Big \{  L_n \big( \tau(\cdot) \big)  + \Lambda_n \big(  \tau(\cdot) \big)  \Big \}  $
			\\
			\midrule
			\textbf{Multitask-Learner} &  \cite{vanderschaar_2017, vanderschaar_2018}  &   $\tau(\bm{x}_i) = \mathbf{f^{\top}} (\bm{x}_i) \mathbf{e} $
			\\
			\midrule
			$\tau$\textbf{-Learner} &  \cite{hahn_2020}  & $\tau (\bm{x}_i)$ as explicit model parameter
			\label{metalearners}
		\end{tabular}
	\end{table}

	\subsubsection{S-Learners}
	
	``Single-Learners", shortened to S-Learners, have been implicitly proposed in two early contributions \citep{hill_2011, foster_2011}, and derive an estimator for CATE by including treatment assignment as ``just another covariate" in the covariate space $\mathcal{X}$, which means that CATE is estimated as
	\begin{equation}
	\tau (\bm{x}_i) = f\big( [\bm{x}_i ~~ 1] \big) - f\big( [\bm{x}_i ~~ 0] \big)~.
	\end{equation}
	An S-Learner fits a single surface $f(\cdot)$, employing the regressors $[\bm{X}_i ~~ Z_i]$, through a base-learner and derives CATE estimates by taking the difference between the two conditional average potential outcomes, which are represented by the fitted $\widehat{f} (\cdot)$ with $Z_i = 1$ and $Z_i = 0$ respectively. The underlying assumption is that the group-specific conditional average potential outcomes stem from the same model, with conditional mean function $f (\cdot)$ and error term $\varepsilon_i$. Regression trees are popular base-learners employed in the context of S-Learners. For instance, \cite{hill_2011} advocates the use of Bayesian Additive Regression Trees (BART), while \cite{foster_2011} of random forests.
	
\begin{figure}[t]
	\centering
	\begin{minipage}{.5\textwidth}
		\centering
		\includegraphics[scale=0.52]{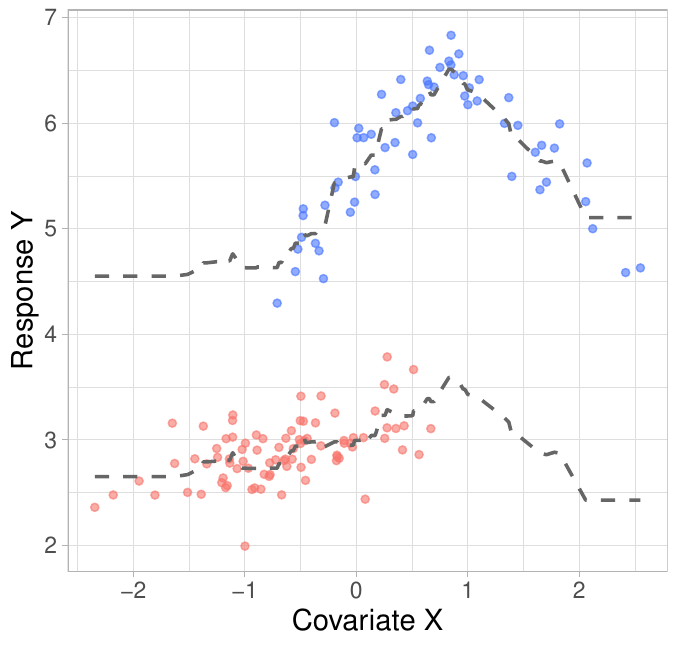}
	\end{minipage}%
	\begin{minipage}{.5\textwidth}
		\centering
		\includegraphics[scale=0.52]{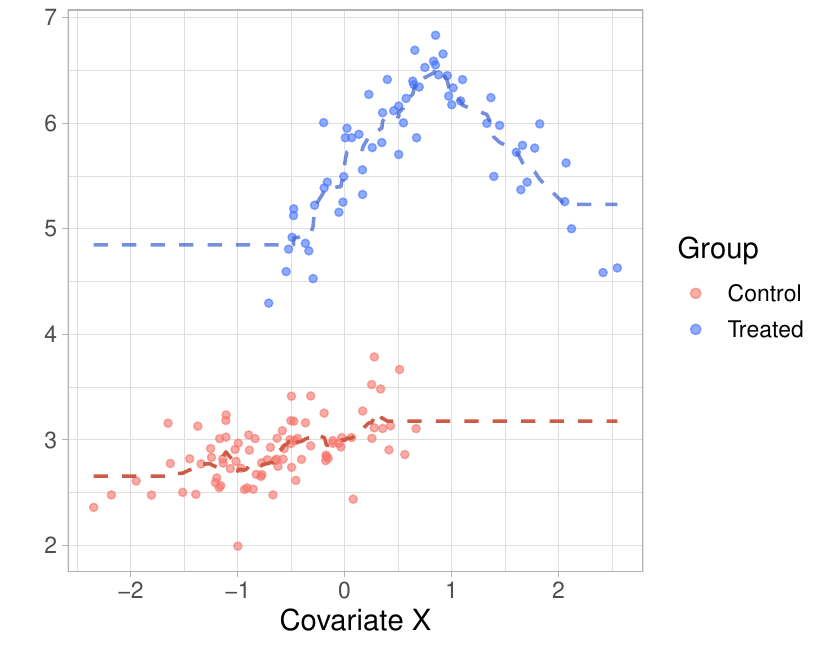}
	\end{minipage}
	\caption{\small Simulated one-covariate data from Section \ref{sec:ProblemSetup}. Left panel: conditional mean fit from a S-Learner BART (dashed grey line). Right panel: conditional mean fit from a T-Learner BART (blue and red dashed lines).}
	\label{STBART}
\end{figure}

	The left panel plot of Figure \ref{STBART} shows a S-Learner BART fit for the conditional mean $\widehat{f} (\cdot)$ of the single-covariate simulated example already encountered in Figure \ref{example}. Notice that the dashed line representing $\widehat{f} (\cdot)$ has a unique color (grey) to emphasize the fact that S-Learner fits a unique surface.

	Since a S-Learner fits a single regression, it is quite restrictive in the way it accounts for the variation in $f(\cdot)$ attributable to $Z$; and it does not take into account the fact that the distribution of the covariates $\bm{X}_i \in \mathcal{X}$ may vary in $Z$, as a result of selection bias. This is obviously more problematic when working with observational data, while it is less so with randomized studies where selection bias is less likely to be present. \cite{vanderschaar_2018} and \cite{hahn_2020} have both identified that the main drawback of S-Learners is their lack of ability in adapting to different levels of sparsity and smoothness across the two treatment groups, since they impose the same regularizing conditions for both treated and control groups. A S-Learner will then perform poorly in a situation where the outcome surface complexity is very different across the two groups. On the contrary, S-Learner is expected to perform well when CATE is of simple form, as the complexity of the conditional average potential outcomes $\mu_Z (\bm{X}_i) = \mathbb{E} \big[ Y_i \mid \bm{X}_i, Z_i \big]$ surfaces does not vary much across treatment groups. For example, consider the case of a S-Learner employing a tree ensemble base-learner, such as BART. Since a tree ensemble method like BART picks splitting variables at each node in each tree randomly, it might not even choose $Z$ as splitting variable in some of the trees in the ensemble, so that $Z$ will possibly be included in most of the trees fitting the response $Y$, but not necessarily in all of them. The exclusion of $Z$ from the splitting rules of a tree in BART is more likely to happen as the number of covariates $\bm{X}_i$ grows larger, in that the model has a larger set of splitting variables to pick from \citep{caron_2021}. This intuitively explains why S-Learners turn out to be appropriate in situations where the complexity of ground-truth CATE is reasonably low, relative to the variation in outcome attributable to the covariates only ($\mathbb{E} \big[ Y_i \mid \bm{X}_i\big]$). It may happen in real world applications, such as clinical studies, that the researcher possesses some domain knowledge regarding the complexity of the treatment effect. As it becomes clearer in the following sections, this is a non-negligible piece of information when it comes to choose a suitable method to perform the heterogeneous treatment effects analysis.

	\subsubsection{T-Learners}

	``Two-Learners'', shortened to T-Learners, derive an estimator for CATE by fitting two separate surfaces for the treated and control groups and computing their difference:
	\begin{equation} \label{eq:tlearner}
	\tau (\bm{x}_i) = f_1(\bm{x}_i) - f_0(\bm{x}_i)~.
	\end{equation}
	Versions of T-Learners can be found in many contributions in the literature. For instance, \cite{athey_2016}, \cite{lu_sadiq_2018} and \cite{powers_2018} offer some examples employing decision trees, random forests and gradient boosted trees as base-learners respectively.
	In contrast to S-Learners, T-Learners separate the two treatment groups when modelling response variable $Y$, and assume that group-specific conditional average potential outcomes are derived from separate regression models, characterized by different conditional means $f_1(\cdot)$ and $f_0(\cdot)$ and independent error terms $\varepsilon_{1i}$ and $\varepsilon_{0i}$. This allows to preserve distributional differences across the two groups that might originate from selection bias, and to take into account different degrees of sparsity and smoothness that vary with $Z$, when regressing $Y$ against $X$. On the other hand, a shortcoming of T-Learners is that, as a result of splitting the sample in two, they do not allow sharing common underlying information between the groups when estimating the two surfaces. This is particularly not ideal in a scenario where individuals in the two groups share the same distributional characteristics, such as in a randomized study.
	
	The right panel plot in Figure \ref{STBART} displays a T-Learner BART fit for the conditional means of the two treatment groups $f_1 (\cdot)$ and $f_0 (\cdot)$, on the same one-covariate simulated example of Figure \ref{example}. Notice that fitted $\widehat{f}_1 (\cdot)$ and $\widehat{f}_0 (\cdot)$ are differentiated by colors (blue and red dashed lines respectively), emphasizing the fact that T-Learner fits two separate models with independent error terms.
	
	T-Learners are expected to work particularly well when complexity of the response surface is very different across treatment groups, and so CATE itself turns out to be a rather complex function. In addition, as formally derived by \cite{vanderschaar_2018}, T-Learners are expected to do well as sample size $N$ goes to infinity, and more observations per group are available to estimate $f_1 (\cdot)$ and $f_0 (\cdot)$. However, this is not usually the case with real world data. For example, in presence of highly imbalanced designs, where one group is larger than the other, splitting the sample in two subgroups might leave too few observations for the estimation of $f_{Z}$ in the smaller group. In the next subsections we will see how this issue is addressed by other Meta-Learners extending the T-Learner framework (X-Learners and Multitask-Learners). On the contrary, T-Learners tend to perform quite poorly in settings where CATE surface is relatively simple and heterogeneity patterns are not so sophisticated, i.e.,~situations where S-Learner usually performs better. Hence if subject-matter knowledge suggests that treatment impact is likely to be significantly complex, a T-Learner might be the preferred choice.

	\subsubsection{X-Learners}  \label{sec:xlearners}
	
	X-Learners have been proposed by \cite{kunzel_2017} as an extension of T-Learners, and derive a CATE estimator in three steps. In the first step, conditional average potential outcomes are fitted as in a T-Learner approach, that is by using two separate regression models for the conditional means  $f_1(\bm{x}_i)$ and $f_0(\bm{x}_i)$, assuming independent error structures. Then in the second step, ``imputed treatment effects" are computed for each group separately; these are defined as the differences between the group-specific observed outcome $Y_i^Z$, and the estimated unobservable conditional average potential outcome $\widehat{Y}_i^{(Z)}$ derived in the first step, more formally:
	\begin{equation}  \label{impTE}
	\begin{split}
	\tilde{D}^1_i & = Y^1_i - \widehat{Y}_i^{(0)} \qquad \text{if} \quad Z_i = 1 \\
	\tilde{D}^0_i & = \widehat{Y}_i^{(1)} - Y^0_i \qquad \text{if} \quad Z_i = 0  ~.
	\end{split}
	\end{equation}
	The second step thus attempts to recover the unobservable differences $ D_i =Y_i^{(1)} - Y_i^{(0)}$ (ITE) separately for the treated and control group by replacing the unobservable potential outcomes with the relative conditional average potential outcome estimates $\widehat{Y}_i^{(1-Z)}$, but using the observed outcome for the other ``actual" outcome $Y_i^Z = Y_i^{(Z)}$, whereas a T-Learner would just use fitted values for both instead.
	In the last step, $\tilde{D}^1_i$ and $\tilde{D}^0_i$ are utilized as response variables in two separate regressions, employing the chosen base-learner (linear regression, random forest, BART, etc.), to obtain estimates of $\hat{\tau}_1(\bm{x}_i)$ and $\hat{\tau}_0(\bm{x}_i)$, using covariates $\bm{X}_i$ as regressors. These two independent regressions can be depicted as
	\begin{equation} \label{lastXstep}
	\begin{split}
	\tilde{D}^1_i & = \tau_1(\bm{X}_i) + \eta_{1i}  \qquad \text{if} \quad Z_i = 1 \\
	\tilde{D}^0_i & = \tau_0(\bm{X}_i) + \eta_{0i}  \qquad \text{if} \quad Z_i = 0   ~ ,
	\end{split}
	\end{equation}
	where the two estimated quantities $\hat{\tau}_1(\bm{x}_i)$ and $\hat{\tau}_0(\bm{x}_i)$ from (\ref{lastXstep}) are group-specific CATE estimates. The final CATE estimate is then obtained through a weighted average of the two group-specific CATE estimates,
	\begin{equation} \label{Xweight}
	\widehat{\tau}(\bm{x}_i) = g(\bm{x}_i) \widehat{\tau}_0(\bm{x}_i) + \big(1 - g(\bm{x}_i) \big) \widehat{\tau}_1(\bm{x}_i)  ~ ,   
	\end{equation}
	where $g(\bm{x}_i) \in [0,1]$ is a given weighting function. The authors propose to set $g(\cdot)$ equal to a propensity score estimate $g(\bm{x}_i) = \widehat{\pi}(\bm{x}_i)$, but this can also take other forms (e.g.~$g(\bm{x}_i) = 1$ or $g(\bm{x}_i) = 0$). 
	
	The intuition behind the last weighting step, that particularly characterizes X-Learners, is the following. When we are in presence of an unbalanced design and we fit a T-Learner, we generally pick a more complex model for the large treatment group and a simpler one for the small treatment group to avoid overfitting. While the model for the larger group is likely to be well-specified, since we observe a lot of data points in that group, the model for the smaller group might not be very representative of the true conditional average potential outcome function $\mu_{Z} (\bm{x}_i) = \mathbb{E} \big[ Y^{(Z_i)} \mid \bm{X}_i \big]$, as we only observe a handful of data points. Nonetheless, the simpler model, which is then also employed to obtain group-specific CATE estimates of the smaller group $\hat{\tau}_Z(\bm{x}_i)$, might be highly representative of the CATE function instead, so that $\hat{\tau}_Z(\bm{x}_i)$ is actually very close to the true $\tau(\bm{x}_i) = \mu_{1} (\bm{x}_i) - \mu_{0} (\bm{x}_i)$.
	
	The choice of $g(\bm{x}_i) = \widehat{\pi}(\bm{x}_i)$ allows to properly assign higher weight to the simpler model's CATE estimates, since e.g.~if the treated group is small, then $\widehat{\pi}(\bm{x}_i)$ will generally be small, and the final CATE estimates $\hat{\tau}(\bm{x}_i)$ will thus be close to $\hat{\tau}_1(\bm{x}_i)$. Choosing $g(\bm{x}_i) \in \{0, 1\}$ is useful in scenarios where the groups are very unbalanced, where more extreme values are necessary to nudge the final CATE estimates $\hat{\tau}(\bm{x}_i)$ towards the smaller group's estimates $\hat{\tau}_Z(\bm{x}_i)$.
	
	For this reason, in unbalanced studies, where T-Learners would yield unnecessarily complex estimates of CATE, X-Learners attempt to improve accuracy by	re-balancing group-specific CATE estimates through propensity score weighting. In this way, they avoid overfitting and revert back to simpler CATE patterns. A final remark about X-Learners, which is naturally valid for all propensity methods, is that careful specification of the propensity model is required to effectively improve precision in CATE estimates through the last balancing step. Poor propensity estimates might not deliver the desired results.

	\begin{figure}[t]
		\centering
		\begin{minipage}{.5\textwidth}
			\centering
			\includegraphics[scale=0.52]{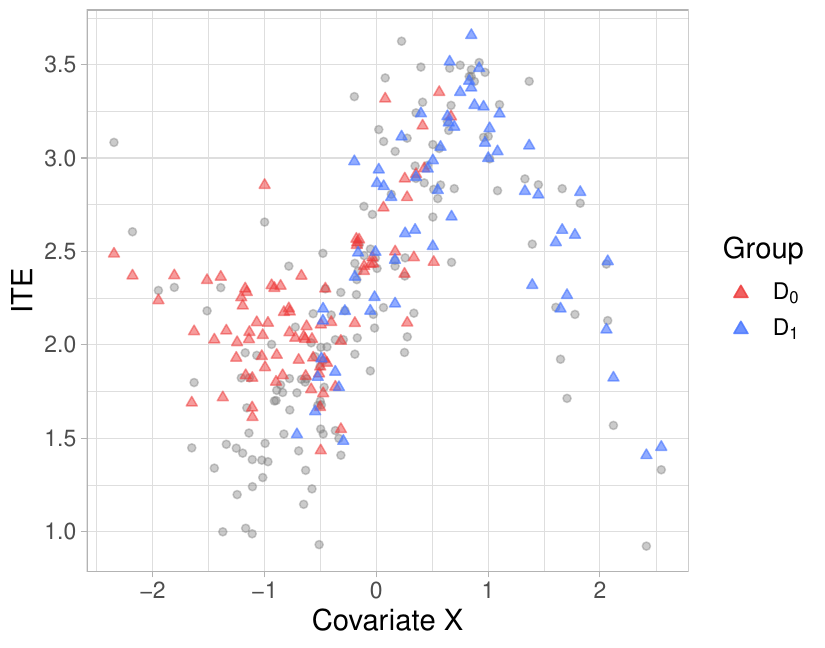}
		\end{minipage}%
		\begin{minipage}{.5\textwidth}
			\centering
			\includegraphics[scale=0.52]{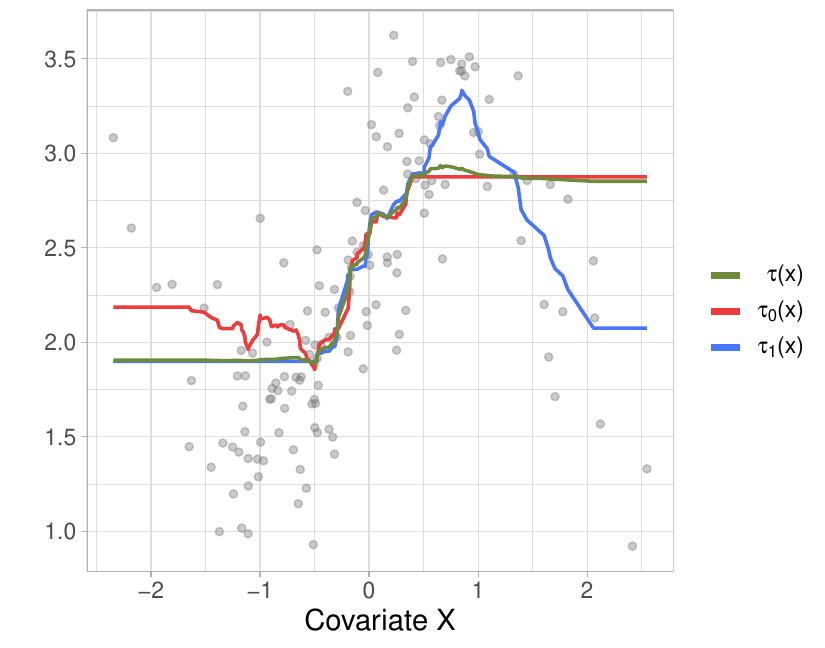}
		\end{minipage}
		\caption{\small X-Learner BART applied to the simulated one-covariate example. Left panel: unobservable ITE (grey dots) and imputed treatment effects $D^1$ and $D^0$ (blue and red triangles), estimated as in (\ref{impTE}) using T-Learner BART. Right panel: group-specific CATE estimates (blue and red dashed lines) obtained from the two regressions in (\ref{lastXstep}), and final weighted CATE estimates (green dashed line) obtained from the re-balancing step in (\ref{Xweight}).}
		\label{XBART}
	\end{figure}

	Figure \ref{XBART} offers a simple example of a X-Learner, with BART as base-learner, applied to the one-covariate simulated data encountered in Figure \ref{example} and Figure \ref{STBART}. X-Learner's first step essentially derives, via T-Learner, the same BART estimates seen in the right panel plot of Figure \ref{STBART}. The output of the second step, namely the imputed treatment effects $\tilde{D}^1_i$ and $\tilde{D}^0_i$, are depicted in the left panel plot of Figure \ref{XBART} (red and blue triangles), together with the true ITE (grey dots). The graph on the right instead shows the estimated group-specific CATE $\hat{\tau}_1(\bm{x}_i)$ (blue dashed line) and $\hat{\tau}_0(\bm{x}_i)$ (red dashed line), derived from the two regressions in (\ref{lastXstep}), and the final CATE estimate $\widehat{\tau}(\bm{x}_i)$ (green dashed line), obtained from the weighting step in (\ref{Xweight}). Propensity score estimates employed for the weighting step were retrieved via probit version of BART. Notice that the final CATE estimates $\widehat{\tau}(\bm{x}_i)$ lie in between the two fitted group-specific $\hat{\tau}_1(\bm{x}_i)$ and $\hat{\tau}_0(\bm{x}_i)$.

	\subsubsection{R-Learners}
	
	R-Learner was proposed by \cite{wager_2019} as a two-stage meta-algorithm, which aims at minimizing a loss function, specifically defined on CATE, through parameter tuning. The derivation of the two-step procedure stems from \cite{robinson_1988} decomposition of the outcome model in (\ref{onestage}). Define the following two quantities:
	\begin{align}  \label{robinson}
	\begin{split}
	Y_i & =  \mu(\bm{X}_i) + \tau(\bm{X}_i) Z_i + \varepsilon_i   \\
	m(\bm{X}_i) & = \mathbb{E} (Y_i \mid \bm{X}_i) = \mu(\bm{X}_i) + \tau(\bm{X}_i) \pi(\bm{X}_i)
	\end{split}
	\end{align}
	as the outcome model and the \emph{conditional mean outcome} model, respectively. Under this setup, unconfoudedness assumption implies that the error term is such that $\mathbb{E} \big[ \varepsilon_i \mid \bm{X}_i, Z_i \big] = 0 $. Notice that under this parametrization $\tau(\bm{X}_i)$ (CATE) enters explicitly in the outcome regression model. By combining the two quantities above, \cite{robinson_1988} noticed that the outcome model can be parametrized as: 
	\begin{equation} \label{transout}
	Y_i - m(\bm{X}_i) = \Big(Z_i - \pi(\bm{X}_i) \Big) \tau (\bm{X}_i) + \varepsilon_i ~ .
	\end{equation}
	Starting from this decomposition, \cite{wager_2019} derive a loss function that can be used for parameter tuning in the estimation of CATE; the optimal CATE estimates are defined as the minimizer of the following loss function:
	\begin{equation} \label{tauargmin}
	\tau(\bm{X}_i) = \argmin_{\tau} \Biggl\{ \mathbb{E} \Bigg[ \Big( \big(  Y_i - m(\bm{X}_i)  \big)   -  \big(  Z_i - \pi(\bm{X}_i)  \big)  \tau(\bm{X}_i)  \Big)^2 \Bigg]  \Biggr\} ~ .
	\end{equation}
	The intuition of eq.\,(\ref{tauargmin}) is the following. Suppose that an individual $i$ is characterized by an extreme propensity value $\pi(\bm{x}_i) \approx 0$; thus its realized treatment assignment is (almost) deterministically $Z_i = 0$. In this extreme scenario, eq.\,(\ref{tauargmin}) boils down to simple Mean Squared Error (MSE) minimization for the conditional mean outcome model $m(\bm{x}_i)$, as in a standard regression problem. Hence, the term $\big( Z_i - \pi(\bm{X}_i) \big) \tau(\bm{X}_i)$ subtracted serves as a de-biasing term that grows larger with the discrepancy between the realized treatment assignment $Z_i$ and the propensity score $\pi(\bm{x}_i)$, and is supposed to tackle selection into treatment imbalance through propensity score re-weighting.
	
	The idea is that a base-learner that relies on parameter tuning (e.g.~random forest or gradient boosted trees) can be tuned on the modified parametrization of the outcome model in (\ref{transout}), which includes a version of the outcome net of the baseline impact of the covariates $\bm{X}_i$ on $Y_i$, $m(\bm{X}_i)$, and propensity score balancing, instead of being tuned on the raw outcome $Y_i$ as one would do in an S- or T-Learner framework.
	Since we cannot observe directly the quantities in (\ref{tauargmin}) for the minimization problem, the R-Learner replaces them with estimates, through the following two-step approach:
	\begin{enumerate} 
		\item Split the data into $k$-folds (5 or 10 suggested). Fit nuisance functions $\widehat{m}(\bm{x}_i)$ and $\widehat{\pi} (\bm{x}_i)$ (on a portion of left-out data) by minimizing usual prediction errors via cross-validation
		
		\item Plug in estimates from the first step to estimate $\hat{\tau} (\bm{x}_i)$, by minimizing the regularized sample equivalent of (\ref{tauargmin}) via parameters tuning on the $k$-folds, that is:
		\begin{equation}   \label{rlearner}
		\begin{split}
		\widehat{\tau}(\bm{X}_i) & = \argmin_{\tau} \Big \{  \widehat{L}_n \big( \widehat{\tau}(\bm{X}_i) \big)  + \Lambda_n \big(  \widehat{\tau}(\bm{X}_i) \big)  \Big \} ~ , \quad \text{where}  \\
		\widehat{L}_n \big( \widehat{\tau}(\bm{X}_i) \big) & = \frac{1}{N} \sum_{i = 1}^{N}  \Bigg(  \Big(  Y_i - \widehat{m}^{-i} (\bm{X}_i)  \Big)  -  \Big( Z_i - \widehat{\pi}^{-i} (\bm{X}_i)  \Big)  \widehat{\tau}(\bm{X}_i)  \Bigg)^2 ~ ,
		\end{split}
		\end{equation}

	\end{enumerate} 
	
	\noindent and where $\Lambda_n \big(  \hat{\tau}(\cdot) \big)$ is a term representing regularization (e.g.~L1 or L2 regularization, splines smoothness penalization, dropout, etc.). The super-script $(-i)$ refers to the $i$-th observation being held-out from the estimation subsample, and used for $k$-fold cross-validation (or even more computationally intense leave-one out cross-validation).
	
	The R-Learner setup resembles, and is in fact inspired by, the doubly robust approach to the estimation of average treatment effects \citep{robinson_1988, duflo_2018}, with the main difference that the second stage in the R-Learner is specifically designed for heterogeneous treatment effects estimation with non-parametric methods \citep{wager_2019, athey_2019_application}. The strength of R-Learners lies in their two-stage procedure, where the first stage takes care of predicting the nuisance functions $\widehat{m}(\bm{x}_i)$ and $\widehat{\pi} (\bm{x}_i)$, while the second focuses on CATE estimation by constructing a direct loss function on it. In this way, R-Learners ensure that regularization is implemented separately for nuisance functions and for CATE. This is intuitively a desirable feature when CATE is of different complexity (in most cases smoother and sparser) compared to the nuisance functions, in that more (or less) aggressive regularization can be conveyed when estimating it, compared to that conveyed in estimating the baseline marginal effect of the covariates on the outcome, $m(\bm{x}_i)$. This particular feature is also shared by $\tau$-Learners, introduced in the next subsection, where direct regularization on CATE is instead applied in the form of Bayesian shrinkage priors \citep{hahn_2020, caron_2021}.
	
	The loss minimization procedure described by the R-Learner framework can generally involve any supervised learning method. The original work \citep{wager_2019} focuses particularly on the use of LASSO and gradient boosted trees, whose parameters are tuned to minimize the R-loss in (\ref{rlearner}). Furthermore, as we will briefly discuss in Section \ref{sec:modelsel}, the R-loss can in principle be used also to evaluate CATE estimates $\hat{\tau}(\cdot)$ derived from different models, or more generally different Meta-Learners. 
	
	A final remark about R-Learners concerns the popular Causal Forest model for CATE estimation \citep{athey_2019, athey_2019_application}, which we are going to include in our comparison of methods based on simulated studies later in Section \ref{sec:simu}. As stated in \cite{athey_2019_application}, Causal Forest can be viewed as a regression forest method motivated by the R-Learner setup. Indeed, its latest implementation uses separate regression forests to fit the nuisance functions and then trains another forests on the CATE loss function in (\ref{rlearner}).

	\subsubsection{Multitask-Learners}
	
	The idea of multitask-learning, or multi-output learning, for causal inference was explicitly introduced by \cite{vanderschaar_2017} and \cite{vanderschaar_2018}, in the context of Gaussian Processes. The multitask perspective on CATE estimation consists in viewing the two potential outcomes $Y_i^{(0)}$ and $Y_i^{(1)}$ as output of a function $f: \mathcal{X} \rightarrow \mathbb{R}^2$, with $d$-dimensional input space and $2$-dimensional output space, where the output space is indexed by $Z_i$, that acts as ``task identifier". CATE estimator is defined as the difference between the elements of the $2$-dimensional output of $f (\cdot)$, i.e.,
	\begin{equation} \label{multitau}
	\widehat{\tau}(\bm{x}_i) = \widehat{f}_1(\bm{x}_i) - \widehat{f}_0(\bm{x}_i) = \mathbf{\hat{f}^{\top}} (\bm{x}_i) \bm{\xi} ~ , \quad ~ \text{where} \quad \bm{\xi^{\top}} = [ -1 ~~ 1 ] ~ .
	\end{equation}
	
	Equation (\ref{multitau}) displays a very similar formulation to a T-Learner; and as in the T-Learner procedure, the sample is practically split into the two subgroups for the estimation. However, the advantage of viewing CATE estimation as a multitask problem is that, instead of estimating the two potential outcomes independently as one would do in a T-Learner or X-Learner, they are estimated ``jointly", through the specification of some hyperparameters that trigger a joint optimization for the two ``tasks": learning $f_{Z=1}$ and $f_{Z=0}$. Hence, this approach fits separate conditional mean functions (as in a T-Learner), but at the same time attempts to recover common underlying patterns between the two groups (as in an S-Learner) that would be otherwise lost due to the sample split. A side advantage of multioutput learning is related to the fact that joint estimation is convenient in cases where a treatment arm features a substantially smaller number of units, in that the process of borrowing information from the larger group becomes beneficial in fitting the conditional average potential outcome of the other group.
	
	In the case of \cite{vanderschaar_2017}, multitask learning is induced through the specification of a particular structure in the stationary kernel function of the Gaussian Process prior. This specific type of GP prior kernel is often known as ``coregionalization" kernel, and it is designed to induce correlation in the estimation of vector-valued functions $\bm{f} (\cdot)$ that map to multiple outcomes; this forces the underlying functions constituting $\bm{f} (\cdot) = [ f_0 ~~ f_1]$ to share similar patterns. We refer to \cite{alvarez_2012} for a more detailed discussion on multioutput learning in GPs. The method is labelled as \emph{Causal Multitask Gaussian Process} (CMGP). \cite{vanderschaar_2018} then proposed a similar method where multitask learning is induced via a non-stationary version of the GP kernel function (\emph{Non-Stationary Gaussian Process} - NSGP).
	
	\cite{vanderschaar_2017} and \cite{vanderschaar_2018} are two example of Multitask-Learners employing Gaussian Process regression as a base-learner, but there are different ways of inducing multitask learning using other types of base-learners (e.g.~linear regression, tree-based methods). In particular, a popular recent contribution that implicitly falls into the Multitask-Learners category is represented by \cite{johansson_2016} and \cite{shalit_2017} implementation of representation learning for CATE estimation (the methods are known with the name of Balancing Neural Network and/or Counterfactual Regression). The idea behind this method is to specify a deep learning model, in the loose form of an encoder structure where a ``balancing" representation of the covariates is learnt by simultaneously minimizing a distance metric between the two distributions of the group-specific representations and a loss function on the fitted conditional average potential outcomes $f_Z(\cdot)$. The goal of the deep neural network structure is to produce counterfactual outputs that generate from an approximation of a randomized study (expressed by the balancing representation). This model can be easily viewed as a form of Multitask-Learner, since the parameters in the deep learning model are shared across the $Z$ tasks (``hard parameter-sharing").
	
	Due to their similarity with T-Learners in deriving a CATE estimator, we expect Multitask-Learners to perform better when complexity of the response surfaces $f_1, f_0$ varies across groups, and CATE itself turns out to be a rather complex function. Nonetheless,  Multitask-Learners avoid extremely complex CATE functions as in a T-Learner setting, via borrowing of information across groups/tasks; and in this way (similarly to X-Learners) they also address the potential issue of unbalanced groups.

	\subsubsection{$\tau$-Learners}  \label{sec:taulearner}
	
	The last type of Meta-Learner presented in this work was developed by \cite{hahn_2020}, under the name of ``Bayesian Causal Forest". The authors tackle the problem of CATE estimation with a Bayesian approach, where they exploit the same parametrization seen in the context of R-Learners. Particularly, they noticed that the parametrization 
	\begin{equation} \label{bcf}
	Y_i = \mu(\bm{X}_i) + \tau (\bm{X}_i) Z_i + \varepsilon_i ~ ,
	\end{equation}
	can be viewed as a Bayesian regression framework where the \emph{prognostic score}, defined as the impact of the covariates $\bm{X}_i \in \mathcal{X}$ on the outcome $Y_i$ in absence of the treatment, $\mu(\bm{x}_i) = \mathbb{E} \big[ Y_i \mid Z_i = 0, \bm{X}_i = \bm{x}_i \big]$, plays the role of the intercept, while $\tau (\bm{x}_i)$ the role of the slope. 
	In this perspective, CATE is an explicit parameter of the model and thus can be treated in a Bayesian fashion through the specification of a prior distribution $p \big( \tau(\cdot) \big)$, which can be shaped to convey prior knowledge and more targeted regularization that can capture even simple patterns of heterogeneity \citep{caron_2021}. Bayesian Causal Forest of \cite{hahn_2020} is composed by a pair of separate and independent BART priors placed on $\mu (\cdot)$ and $\tau (\cdot)$ respectively, but the parametrization in (\ref{bcf}) can be exploited using other Bayesian regression methods (e.g.~Gaussian Process, Dirichlet Process regression, etc.).
	
	In addition to the parametrization shown in (\ref{bcf}), \cite{hahn_2020} make use of the two-stage procedure seen in (\ref{twostage}), Section \ref{sec:ProblemSetup}. The two-stage approach is motivated by the presence of a particular type of confounding, which the authors in \cite{hahn_2018} and \cite{hahn_2020} call \emph{Regularization Induced Confounding} (RIC). The intuition behind RIC is the following: regularization applied directly on the two curves $f_1, f_0$ featuring in a T-Learner (eq.\,(\ref{eq:tlearner})) may have unintended consequences on the induced regularization on $\tau (\cdot)$, leading to biased estimates of CATE. RIC is shown to have a stronger effect when there is strong confounding, such as in presence of \emph{targeted selection}, that is when individuals are selected into treatment based on the prediction of otherwise adverse outcome. Targeted selection implies a potential strictly monotone relationship between the propensity score $\pi (\bm{x}_i)$ and the prognostic score $\mu(\bm{x}_i) = \mathbb{E} \big[ Y_i \mid \bm{X}_i, Z_i = 0 \big]$, and is rather common in studies of observational nature. The proposed way to tackle confounding from targeted selection is precisely to use the two-stage representation illustrated in (\ref{twostage}), where a probabilistic estimate of the propensity score $\widehat{\pi} (\bm{x}_i)$, obtained from the first stage regression, is added to the covariates for the estimation of $\mu(\bm{x}_i) = \mathbb{E} \big[ Y_i \mid \bm{X}_i, Z_i = 0 \big]$ in the second stage, to account for their potential relationship.  
	
	We name the above approach $\tau$-Learner, as it involves an explicit parametrization in terms of $\tau (\bm{x}_i)$ (shared with R-Learners) and a direct Bayesian approach to CATE estimation. \cite{hahn_2020} specifically make use of BART for estimation of $\mu(\bm{x}_i)$ and $\tau(\bm{x}_i)$, but any other Bayesian method could potentially work. 
	
	As a further advantage, the direct Bayesian approach returns full predictive posterior distribution on CATE, which conveniently allows the computation of point estimates as well as credible intervals. This feature is shared also by Bayesian implementation of S-Learners \citep{hill_2011} and can be usefully employed to check for causal common support, as showed by \cite{hill_2013}. On the contrary, frequentist implementations easily provide point and confidence interval estimates, but cannot return full predictive distribution without techniques such as jackknife or bootstrapping.

	\subsection{Model Selection} \label{sec:modelsel}
	
	Model selection is a challenging problem in causal inference, the main reason being that researchers cannot access counterfactual outcomes $Y_i^{(1-Z_i)}$ and thus observe the difference $Y_i^{(1)} - Y_i^{(0)}$, $\forall i \in \{ 1,..., N \}$, which distinguishes it from other classical model selection problems. The aim here is to ideally select a model $\mathcal{M}^* \in \{ \mathcal{M}_1, ..., \mathcal{M}_d \}$, which minimizes a loss function on the estimated CATE $\hat{\tau} (\bm{x}_i)$. CATE squared-loss function, of the type $\mathbb{\ell}( \hat{\tau} , \tau ) = \mathbb{E} [(\hat{\tau} - \tau)^2] $, is referred to as \textit{Precision in Estimating Heterogeneous Treatment Effects} (PEHE) \citep{hill_2011} and takes the following form:
	\begin{equation}  \label{PEHE}
	\mathbb{E} \left[ \big( \hat{\tau} (\bm{x}_i) - \tau (\bm{x}_i) \big)^2 \mid \bm{X}_i = \bm{x}_i \right] ~ .
	\end{equation}
	PEHE would be the ideal loss function to use, but cannot be computed because of partial observability of the POs. Moreover, typical loss functions for standard regression problems, such as prediction error evaluated on the outcome $Y_i$ for estimating conditional average POs, are not reliable measures for the goodness of the final CATE estimates. For example, as discussed in Section \ref{sec:xlearners} in the context of X-Learners, fitting good POs models in terms of their prediction error is not sufficient to guarantee good CATE estimates in unbalanced designs. Attempts have been made in the literature to render the estimation of PEHE feasible, through plug-in estimates of $\tau_i$, but to our knowledge none of them have been successful and commonly used so far \citep{tibshirani_2018}. Finally, the R-loss encountered in (\ref{rlearner}) in the R-Learner framework can be in principle utilized to evaluate CATE estimates $\hat{\tau}(\cdot)$ stemming from any other Meta-Learning framework. However, this implicitly entails assuming \cite{robinson_1988}'s, and thus R-Learner's, parametrization $Y_i = \mu(\bm{x}_i) + \tau (\bm{x}_i) Z_i + \varepsilon_i$, with common error term across groups.

	From a more practical perspective, the problem of model selection for CATE inference can be decomposed into three main tasks:
	
	\begin{itemize}
	
	\item[1)] \textbf{Causally sufficient variable selection}. By causally sufficient variable selection we indicate a step which is aimed at ideally partitioning covariates into four distinct categories, namely: i) confounders, i.e.~common causes of $Z_i$ and $Y_i$, to be included in both outcome and propensity model; ii) predictors of $Z_i$, to be included in the propensity model only; iii) predictors of $Y_i$, to be included in the outcome model only; iv) moderators of the treatment effects, which are a (not necessarily strict) subset of the outcome's predictors entering the CATE model only. In the case of ``direct methods" not relying on propensity score adjustment, the problem naturally reduces to the specification of the outcome model only. The ``causally sufficient" terminology here relates to the inclusion of confounders, which represents the smallest set of covariates to condition on that guarantees unbiased CATE estimates, while variable selection in propensity and outcome model is meant to improve estimates' precision instead. Naturally, in empirical applications with large datasets, manual variable selection is not feasible, so one typically resorts to regularization techniques, after assuming unconfoundedness (i.e.~we observe and include all confounders in the model). The interesting sub-task in heterogenous treatment effects estimation is that of detecting the main moderators, possibly amongst several covariates. As we will discuss in detail in Section \ref{sec:realdata}, R- and $\tau$-Learners have the comparative advantage to other Meta-Learners that they provide a straightforward framework that allow this directly \citep{caron_2021}. As also described earlier, by exploiting \cite{robinson_1988} parametrization they specify a direct regularized model on CATE, that can easily return interpretable measures of variable (in this case, moderators) importance. For example, a LASSO regression implementation of R-Learner \citep{wager_2019} would return a sparse vector of coefficients for the moderators; a shrinkage prior implementation of Bayesian Causal Forest can return posterior splitting probabilities \citep{caron_2021}.   
	
	\item[2)] \textbf{Base-learner selection} refers to the problem of finding the best regression algorithm for fitting the surfaces of interest via the outcome and propensity models. This problem similarly arises in standard supervised regression settings. A first step might be related to determining whether a parametric model is sufficient for adequately approximating relationships in the data. Non-parametric regression models provide flexibility to capture more complex patterns from larger samples. Among non-parametric models, specific models might be selected based on the type of regularization techniques they provide. For example, one might consider splines or Gaussian Processes as more appropriate than tree-based methods for certain type of data, as they are better suited for fitting smooth functions. Some Meta-Learners, namely T-, X- and R-Learners offer the opportunity of employing more than just one base-learner. Within the T- and X-Learners frameworks, different learners can be used for the treated and the control group. For example, if the treated group has very few instances compared to the control, a linear model is likely to be more appropriate. Similarly, in R-Learners, different models can be adopted for fitting $m(\bm{x}_i)$ and then $\tau(\bm{x}_i)$.

	\item[3)] \textbf{Meta-Learner selection}. Finally, the chosen base-learner has to be paired to one of the Meta-Learners subroutines described in earlier subsections. Note that some of the Meta-Learners presented above do not support all types of base-learners. While S-, T-, X- and R-Learners allow for a high degree of flexibility in the choice of a base-learners, the other frameworks are a bit more selective. $\tau$-Learners envisage the use of Bayesian inference and non plug-in methods, and have been currently implemented in the context of BART \citep{hahn_2020, caron_2021}. Multitask-Learners only allow for multi-output learning algorithms (e.g~BART have not been extended to multi-output problems yet). As we will discuss in more detail in the next section, the choice of a Meta-Learner is primarily based on domain knowledge about the study at hand, by recognising study-specific characteristics (i.e.~suspected complexity of heterogeneity patterns, treatment groups imbalance, etc.), and linking them to the properties of meta-learning framework. 
	\end{itemize}

	\begin{table}
		\caption{Tested models}
		\label{testedmodels}
		\small
		\centering
		\begin{tabular}{lcl}
			\toprule
			\textbf{Meta-Learner}     &  \textbf{Label}     &  \textbf{Base-Learner}  \\
			\midrule
			
			\multirow{2}{3cm}{S-Learners}  &  \multirow{1}{3.5cm}{\centering S-RF}  &   Random Forest  \\
			&  S-BART  &   BART \\
			\midrule
			
			\multirow{2}{3cm}{T-Learners}  &  \multirow{1}{3.5cm}{\centering T-RF}  &   Random Forest  \\
			&  T-BART  &   BART \\
			\midrule
			
			\multirow{2}{3cm}{X-Learners}  &  \multirow{1}{3.5cm}{\centering X-RF}  &   Random Forest  \\
			&  X-BART  &   BART \\
			\midrule

            \multirow{3}{3cm}{R-Learners}  &  \multirow{1}{3.5cm}{\centering R-LASSO}  &   LASSO Regression  \\
			&  R-BOOST  &   Gradient Boosted Trees \\
			&  CF  &   Causal Forest \\
			\midrule
			
			\multirow{2}{3cm}{Multitask-Learners}  &  \multirow{1}{3.5cm}{\centering CMGP}  &   Causal Multi-task GP (Multioutput GP) \\
			&  NSGP  &   Non-Stationary GP (Multioutput GP) \\
			\midrule
			
			\multirow{1}{3cm}{$\tau$-Learners}  &  \multirow{1}{3.5cm}{\centering BCF}  &   Bayesian Causal Forests (BART)  \\
			
			\bottomrule
		\end{tabular}
	\end{table}

	\section{Simulation studies} \label{sec:simu}
	
 	In this section we report and comment on results from two different semi-simulated studies, carried out to compare performance of some of the models presented above in estimating CATE. A third supplemental semi-simulated study can be found in the Appendix Section \ref{sec:appendix}. A semi-simulated study here consists in simulating only the outcome surface $Y_i$ in the tuple $\mathcal{D}_i = \{ \bm{X}_i, Z_i, Y_i \} $, starting from real-world $\bm{X}_i$ and $Z_i$. In the case of observational semi-simulations, \cite{hill_2011} introduced a practical way of recreating an observational study from a randomized one. This is essentially done by leaving out a non-random portion of the treated group, so that treatment assignment is no longer randomized. Recreating an observational study from a purely randomized one has the main advantage of ensuring control over the selection mechanism, such that common support is guaranteed to hold, in this case, at least for the treated group. This means that average treatment effects on the treated (ATT and CATT) are identifiable, while those on the controls (ATC and CATC) are not.
	
	We provide results based on the analysis of two real world randomized controlled trials, after transforming them into observational studies. The first semi-simulated setup employs the IHDP dataset, firstly introduced by \cite{hill_2011} and popular in both the computer science and statistics literature on CATE estimation. The second and third setups instead employ the ACTG-175 dataset, and differ in the way the outcome and CATE are generated. Both code and the datasets to reproduce the results in this section are publicly available\footnote{\url{https://github.com/albicaron/EstITE}.}. As mentioned above, we present here below results from the IHDP data simulation and one of the two setups involving the ACTG-175 data, while we leave the other ACTG-175 setup in the Appendix, Section \ref{sec:appendix}. 
	
	The models we test are the following: random forests and BART respectively as S-, T- and X-Learners; LASSO regression, gradient boosted trees as R-Learners, and Causal Forest, which is a particular implementation of random forests as an R-Learner \citep{athey_2019, athey_2019_application}; two Multitask-Learners in the form of Multioutput Gaussian Processes, one with stationary kernel (Causal Multitask GP - CMGP) and the other with non-stationary kernel (Non-Stationary GP - NSGP), developed by \cite{vanderschaar_2017} and \cite{vanderschaar_2018} respectively; finally, Bayesian Causal Forest \citep{hahn_2020, caron_2021}, which is a specific implementation of $\tau$-Learner employing BART. A summary of the tested models is provided in Table \ref{testedmodels}.
	
	For each of the two datasets analyzed, in order to provide a comparison of the methods presented above, we computed $\sqrt{\mbox{PEHE}}$ estimates for each of the $B=1000$ Monte Carlo simulations, and we averaged it over all the simulations. Consistently with what we discussed earlier, $\sqrt{\mbox{PEHE}}$ was evaluated only in the covariate space regions corresponding to the treated units (thus on CATT), as overlap is not guaranteed to hold on the covariate space regions of the controls (i.e.~on CATC). Estimates of $\mbox{PEHE}$ were obtained through its sample equivalent, namely:
	\begin{equation}
	\mbox{P}\widehat{\mbox{EH}}\mbox{E}_{\tau} = \frac{1}{N_T} \sum_{i=1}^{N_T} \Big(  \tau(\bm{x}_i) - \widehat{\tau} (\bm{x}_i) \Big)^2 ~ ,
	\end{equation}
	where $N_T$ is the size of the treated group and $\widehat{\tau} (\bm{x}_i)$ is the CATT estimate obtained under the given method, while $\tau(\bm{x}_i)$ is the ground-truth CATT, always unknown in the real world. Data are randomly split in 70\% train set used to train the models, and 30\% test set to evaluate the model on unseen data. $\sqrt{\mbox{PEHE}}$ is reported for both train and test data. Supplementary results on all the simulated experiments regarding bias and $\sqrt{\mbox{PEHE}}$, evaluated also on CATC (out-of-overlap) regions, are provided in the appendix.

	\DIFaddend \begin{figure}[t]
		\centering
		\begin{minipage}{.5\textwidth}
			\centering
			\includegraphics[width=1\linewidth]{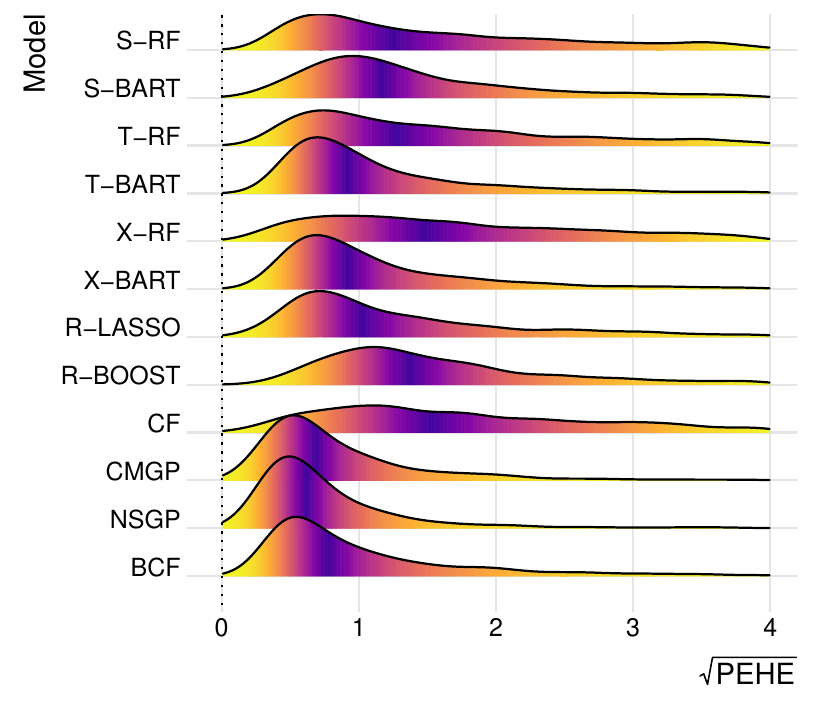}
		\end{minipage}%
		\begin{minipage}{.5\textwidth}
			\centering
			\includegraphics[width=1\linewidth]{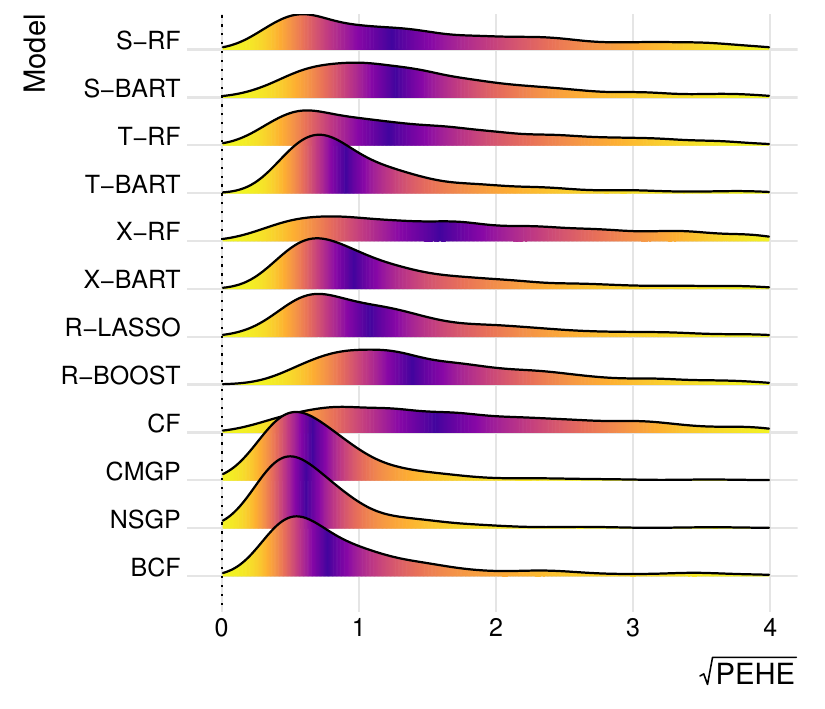}
		\end{minipage}
		\caption{\small $\sqrt{\mbox{PEHE}}$ distribution in the train set (left) and test set (right), IHDP data.}
		\label{fig:IHDPdistr}
	\end{figure}

	\subsection{IHDP data}  \label{sec:IHDP}
	
	The first semi-simulated setup makes use of the IHDP dataset, popular in the literature for CATE estimation and  used for the first time in \cite{hill_2011}. It includes data taken from the Infant Health and Development Program (IHDP), a randomized controlled trial carried out in 1985, aimed at improving health and cognitive status of premature infants with low weight at birth, through pediatric follow-ups and parent support groups. The dataset includes 25 covariates, 6 continuous and 19 binary. The data are transformed into observational by leaving out a non-random portion of the treated individuals, namely those with non-white mothers. This leaves 139 observations in the treated group and 608 in the control group, for a total of 747 observations.

	ITE is derived as the difference between the simulated potential outcomes, which are generated as: 
	\begin{equation} \label{hillPO}
	\begin{split}
	Y^{(0)} & \sim \mathcal{N} \Big( \exp \big( (X+W) \beta_B \big), 1 \Big) ~ ,  \\
	Y^{(1)} & \sim \mathcal{N} \big( X \beta_B - \omega^b_B, 1 \big)  ~ ,
	\end{split}
	\end{equation}
	where $W$ is an offset matrix of same dimension as $X$ with every entry equal to 0.5, and $\beta_B$ is a 25-dimensional vector of regression coefficients $\big( 0, 0.1, 0.2, 0.3, 0.4 \big)$, sampled in each replication $b$ of the experiment with probabilities $ \big( 0.6, 0.1, \allowbreak 0.1, 0.1, 0.1 \big)$ (``Response surface B'' in \cite{hill_2011}). 
	For each replication $b \in \{ 1, ..., 1000 \}$, $\omega^b_B$ is an offset chosen to guarantee that $ATT = \mathbb{E} \big[ Y^{(1)} - Y^{(0)} \mid Z = 1 \big] = 4$.

	\begin{table}
		\caption{\small IHDP and ACTG-175 data. $\sqrt{\mbox{PEHE}_{\tau}}$ estimates $\pm$ 95\% confidence intervals for each tested model on CATT, on train and test sets respectively.}
		\label{simuresults}
		\footnotesize
		\centering
		\begin{tabular}{ccc c cc}
			\toprule
			&   \multicolumn{2}{c}{\textbf{IHDP}} & & \multicolumn{2}{c}{\textbf{ACTG-175}}  \\
			
			\cmidrule(r){2-3}  \cmidrule(r){5-6}
			
			& Train  &  Test &   & Train  &  Test  \\
			
			\midrule
			
			S-RF  		&	3.02 $\pm$ 0.23		&  2.96  $\pm$ 0.25  &   &  0.50 $\pm$ 0.01	&   0.48 $\pm$ 0.01  \\
			S-BART  		&	1.75 $\pm$ 0.11 	&  2.02 $\pm$ 0.15      & 	&  0.43 $\pm$ 0.01	&  0.45 $\pm$ 0.01  \\

			\midrule
			
			T-RF  		&	2.39 $\pm$ 0.17		&  2.28  $\pm$ 0.18  &   &  0.51 $\pm$ 0.01	&   0.49 $\pm$ 0.01  \\
			T-BART  		&	1.35 $\pm$ 0.09 	&  1.31 $\pm$ 0.09      & 	&  0.48 $\pm$ 0.01	&  0.51 $\pm$ 0.01  \\
			
			\midrule
			
			X-RF  		&	3.04 $\pm$ 0.21		&  3.15  $\pm$ 0.24  &   &  \textbf{0.34 $\pm$ 0.01}	&   \textbf{0.36 $\pm$ 0.01}  \\
			X-BART  		&	1.34 $\pm$ 0.09 	&  1.50 $\pm$ 0.12      & 	&  0.44 $\pm$ 0.01	&  0.47 $\pm$ 0.01  \\
			
			\midrule

			R-LASSO  		&	1.78 $\pm$ 0.13 	&  1.82 $\pm$ 0.14      &    &  0.60 $\pm$ 0.01	&   0.63 $\pm$ 0.01	\\
			R-BOOST  		&	2.04 $\pm$ 0.12 	&  2.22 $\pm$ 0.15      &    &  0.47 $\pm$ 0.01	&   0.48 $\pm$ 0.01	\\
			CF  		&	2.88 $\pm$ 0.19 	&  2.84 $\pm$ 0.21      &    &  0.40 $\pm$ 0.01	&   0.39 $\pm$ 0.01	\\
            
            \midrule
            
            CMGP  		&   \textbf{0.89 $\pm$ 0.05} 	&	\textbf{0.84 $\pm$ 0.07}      &	   &  0.42 $\pm$ 0.01 	&	0.43 $\pm$ 0.01	\\
			NSGP		&  	\textbf{0.80 $\pm$ 0.05} 	&	\textbf{0.81 $\pm$ 0.07}      &  	&  0.41 $\pm$ 0.01 	&	0.42 $\pm$ 0.01	\\
			
			\midrule
			
			BCF  		&	1.26 $\pm$ 0.09 	&  1.22 $\pm$ 0.09      &    &  \textbf{0.36 $\pm$ 0.01}	&   \textbf{0.38 $\pm$ 0.01}	\\
			
			\bottomrule
		\end{tabular}  
	\end{table} 	
	
	Given the features of this specific simulated experiment, we might anticipate some of the Meta-Learners' behaviors, based on the properties that we laid out in the previous section. First of all, we notice that, by the way POs are generated, CATE is bound to be a rather complex function. We expect this feature to particularly favours T-Learners and their extensions Multitask-Learners, over S-, X-, R- and $\tau$-Learners, as they tackle CATE estimation problem by fitting two separate functions $f_0$ and $f_1$, which allows to capture very distinct, group-specific, degrees of complexity. Secondly, at a higher base-learner selection level, the conditional average potential outcomes generated in (\ref{hillPO}) are very smooth functions. This implies that base-learners enforcing a higher degree of smoothness via regularization (e.g.~splines, Gaussian Processes, etc.) are well suited for the problem at hand.
	
	Simulation results on performance are reported in Table \ref{simuresults} and Figure \ref{fig:IHDPdistr}. As anticipated by the above considerations, the best models appear to be the multitask Gaussian Processes (CMGP and NSGP) of \cite{vanderschaar_2017} and \cite{vanderschaar_2018}. Also, T-Learners generally display better performance than their S- and X-Learner counterparts (particularly over S-RF, S-BART and X-RF, while X-BART has comparable performance to T-BART). Less anticipated is the performance of BCF ($\tau$-Learner), which comes in second after the GPs. This highlights BCF's ability to convey targeted Bayesian shrinkage on CATE, that evidently allows also to adjust to more (or less) complex CATE surfaces. In Figure \ref{fig:IHDPdistr}, we report the empirical distribution of $\sqrt{\mbox{PEHE}_{\tau}}$ over the $B=1000$ replications on both train and test data, for each of the models. We also notice that tree-based methods are relatively more prone to overfitting.

	An important remark about the data generating process described by (\ref{hillPO}) is that it does not really induce strong confounding, since it is easy for a non-parametric base-learner to distinguish the two underlying polynomials $\mathbb{E} [ Y^{(Z_i)} \mid \bm{X}_i = \bm{x}_i ]$. And since the two polynomials $\mathbb{E} [ Y^{(Z_i)} \mid \bm{X}_i = \bm{x}_i ]$ are extremely different from each other, CATE ends up being an unrealistically complex function. Besides, the fact that noise around $\mathbb{E} [ Y^{(Z_i)} \mid \bm{X}_i = \bm{x}_i ]$ is independently simulated for the two potential outcomes produces extra noise around CATE, $\mathbb{V}( Y_i^{(1)} - Y_i^{(0)} ) = \mathbb{V}(\varepsilon_{i,1}) + \mathbb{V}(\varepsilon_{i,0})$, that renders estimation challenging for every model in general. This implies that a relatively higher number of Monte Carlo replications of the experiment are needed to obtain estimates of $\sqrt{\mbox{PEHE}}$ with sufficiently low variance to effectively compare methods' performance (in this case $B=1000$ appears to suffice). In the ACTG-175 simulated example illustrated in the next section, we will follow the parametrization found in \cite{robinson_1988}, \cite{wager_2019} and \cite{hahn_2020} in the data generating process of the outcome surface, in order to induce stronger confounding (which is believed to be common in observational studies), generate a relatively simpler and more realistic CATE function, and avoid creating unnecessarily high noise around CATE.

	\subsection{ACTG-175 data} \label{sec:ACTG}
	
	The second semi-simulated setup presented here is re-created using the ACTG-175 dataset. The data come from a randomized placebo-controlled trial aimed at comparing monotherapy versus a combination of therapies in HIV-1-infected subjects with CD4 cell counts between 200 and 500 (\cite{doi:10.1056/NEJM199610103351501} for details). As in the case of IHDP data, an observational study is recreated by throwing away a non-random subset of patients, namely those not showing symptomatic HIV infection. The final dataset consists of 813 observations and 12 variables (3 continuous and 9 binary). The list of covariates included in the dataset are shown in Table \ref{AIDSvar}.
	
	\begin{table}
		\caption{ACTG 175 data variables}
		\centering \footnotesize
		\begin{tabularx}{10.5cm}{l | X}
			
			\textbf{Variable}  &  \textbf{Description} \\
			\midrule
			
			\textit{age} &  Numeric	\\	
			\textit{wtkg} &	Numeric	\\	
			\textit{hemo} &	 Binary (hemophilia = 1)	\\	
			\textit{homo} &	 Binary (homosexual = 1)	\\
			\textit{drugs} & Binary (intravenous drug use = 1)	\\
			\textit{oprior} & Binary (non-zidovudine antiretroviral therapy prior to initiation of study treatment = 1) \\
			\textit{z30} &	Binary (zidovudine use in the 30 days prior to treatment initiation = 1)	\\
			\textit{preanti} &	Numeric (number of days of previously received antiretroviral therapy)	\\	
			\textit{race} &	 Binary	\\	
			\textit{gender} &	Binary 	\\
			\textit{str2} &	 Binary: antiretroviral history (0 = naive, 1 = experienced)   \\
			\textit{karnof\_hi} &	Binary: Karnofsky score (0 = $<100$, 1 = $100$)	
			\label{AIDSvar}
		\end{tabularx}
	\end{table}
	
\begin{figure}[t]
	\centering
	\begin{minipage}{.5\textwidth}
		\centering
		\includegraphics[width=1\linewidth]{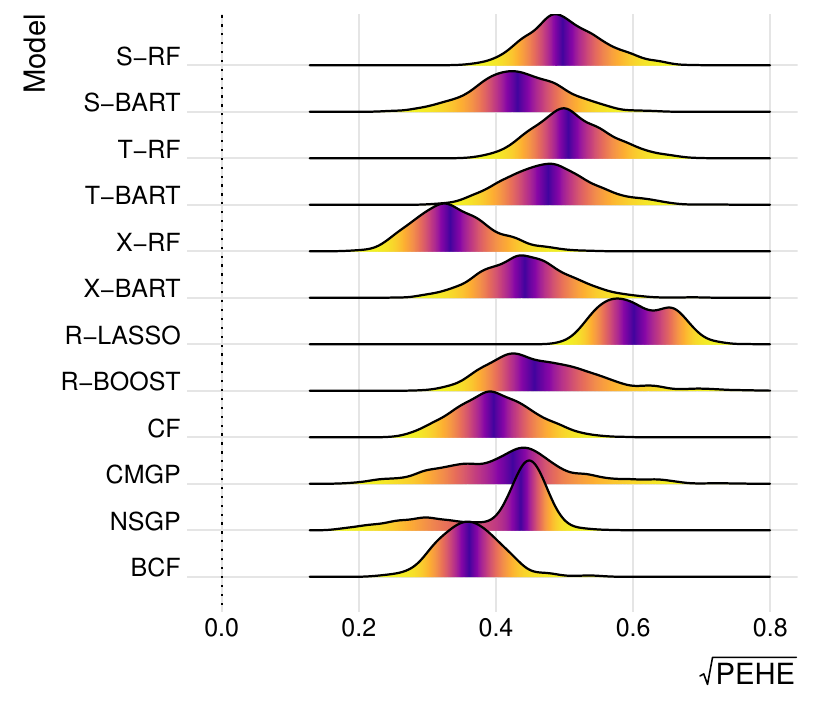}
	\end{minipage}%
	\begin{minipage}{.5\textwidth}
		\centering
		\includegraphics[width=1\linewidth]{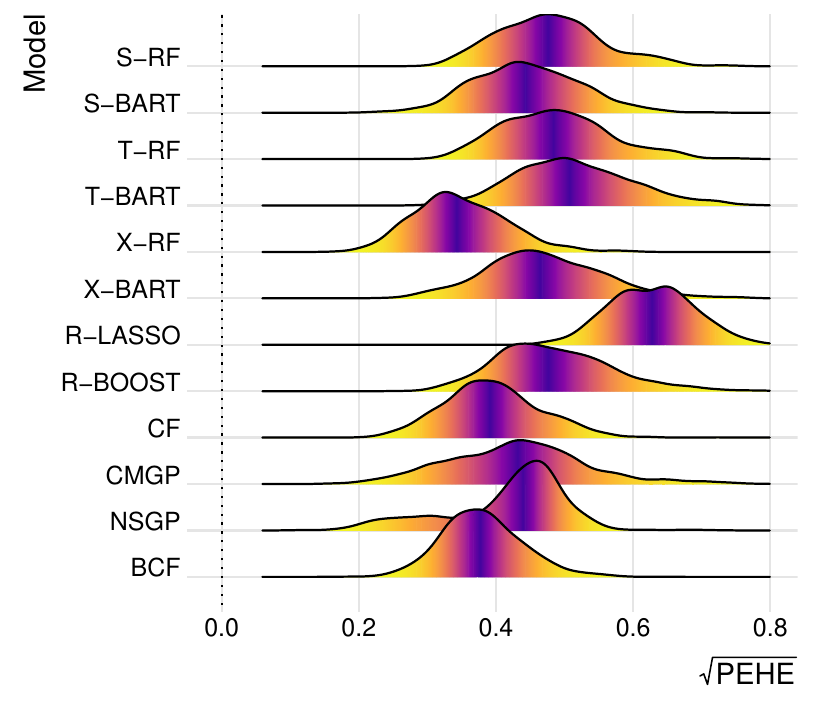}
	\end{minipage}
	\caption{\small $\sqrt{\mbox{PEHE}}$ distribution in the train set (left) and test set (right), ACTG-175 data.}
	\label{fig:AIDSdistr}
\end{figure}

	Unlike the case of IHDP data, response surface $Y_i$ is not generated via simulation of the two potential outcomes. Instead, we generate continuous outcome $Y_i$ according to the parametrization
	\begin{equation}
	Y_i = \mu(\bm{X}_i) + \tau(\bm{X}_i) Z_i + \varepsilon_i ~ ,
	\end{equation}
	which allows to specify CATE directly, instead of starting from the simulation of potential outcomes, and features a single error term $\varepsilon_i$. The prognostic score $\mu(\bm{x}_i)$ and CATE $\tau(\bm{x}_i)$ are generated as:
	\begin{equation}  \label{setup1}
	\begin{split}
	\mu(\bm{x}_i) = & ~ 8 - 0.5hemo - | wtkg - 1 | + 0.5gender - \frac{0.2}{age + 2} \\ 
	& + 0.5karnof_{hi} - 0.5z30  - 0.5race  
	\\[8pt]
	\tau(\bm{x}_i) = & ~ 1 + 0.2wtkg + 2 \phi_Z (wtkg) \cdot (karnof_{hi} + 2) ~ ,
	\end{split}
	\end{equation}
	where $\phi_Z (\cdot)$ is the density of a standard normal distribution. Noise is added by simulating normally distributed i.i.d.~errors $\varepsilon_i \sim \mathcal{N} (0, \sigma^2)$, with homoskedastic standard deviation equal to $\sigma =   0.2(\mu_{max} - \mu_{min})$, where $\mu_{max}$ is the sample maximum of the generated prognostic score, while $\mu_{min}$ is the sample minimum. Notice that, unlike the case of IHDP data, the error term is not simulated independently for the two groups, which avoids imposing too much noise around CATE. This translates into generally lower and less variable $\mbox{PEHE}$ estimates to evaluate the models, compared to the case of IHDP data, as shown both in Table \ref{simuresults} and Figure \ref{fig:AIDSdistr}. As in the IHDP simulated example, $\sqrt{\mbox{PEHE}}$ is evaluated on the treated group only, given that only CATT is guaranteed to be identifiable.
	
	In this second simulated setting, CATE is of rather simple form. Hence, contrary to the IHDP setup, we expect learners that better accommodate simpler CATE functions, such as S-, X- and $\tau$-Learners, to and perform better than T- and Multitask-Learners counterparts. In addition, the design is slightly unbalanced, with 281 individuals in the treated group and 532 in the control, a feature that might favour X-Learners. By inspecting the results reported in Table \ref{simuresults}, we notice that X-RF and BCF are comparably the two best performing methods. As we have pinpointed earlier, this is thanks to their ability to detect simple heterogeneity patterns. S- and X-Learner implementation of BART are then relatively better than T-BART, while S-RF and T-RF do not exhibit any significant difference. CF (random forest R-Learner), which shares the characteristics of conveying targeted regularization with BCF, trails just behind X-RF and BCF. Finally, the two causal multitask GPs perform reasonably well considering that the setup is not favourable to T-type of learners, as they tackle conditional average potential outcomes estimation jointly. Figure \ref{fig:AIDSdistr} depicts again the distribution of $\sqrt{\mbox{PEHE}}$ over the $B=1000$ replications, for both train and test data, for all the tested models.
	
	We employ the ACTG-175 data also in a third semi-simulated setup featuring more complex $\mu(\bm{x}_i)$ and $\tau(\bm{x}_i)$ surfaces, compared to the ones in (\ref{setup1}). Description and results of this third example are provided in the Appendix Section \ref{sec:appendix}.

	\subsection{Recommendations}
	
	We conclude this simulated experiments section with a few recommendations based on the above results. The first regards selection of a base-learner.  One might resort to naive prediction error (e.g.~MSE, MAE, AUC, etc.) to select a base-learner that best fit the baseline relationship between outcome $Y_i$ and covariates $\bm{X}_i$ (which is utilized in R-Learners), or the group-specific conditional average potential outcomes (in T-Learners); or might resort to a simple parametric model if one of the treatment groups has very few data points, as in the Figure 1 example of \cite{kunzel_2017}. Plain classification error can be used instead for selecting a propensity score learner, in the case of propensity models. Notice that vanilla prediction error on the conditional average potential outcomes model is not a proper loss function for the CATE estimation problem per se; thus tuning a T-Learner based on e.g.~MSE minimization does not necessarily return a good CATE model, as highlighted in the context of X-Learners (Section \ref{sec:xlearners}), but it nonetheless helps at least in the choice of a base-learner.
	
	An additional remark regards Bayesian base-learners specifically. As extensively demonstrated by \cite{hill_2013}, Bayesian (non-parametric) regression models offer a natural setup to easily inspect lack of overlap regions and deal with ``out-of-overlap" CATE predictions in some Meta-Learner frameworks. This is helpful since checking whether, and for which specific portion of data, common support holds, can be difficult in high-dimensional settings with a large number of covariates; nonetheless, Bayesian models such as BART or Gaussian Processes allows this rather straightforwardly, at least in those meta-learning setups with single unique error structure, i.e.~S-, R- and $\tau$-Learners. This is because they naturally produce uncertainty measures in the form of predictive posterior distribution estimates on CATE, which tend to be large when one tries to make out-of-overlap predictions.
	
	As for recommendations about the choice of a Meta-Learner, we have observed that Bayesian Causal Forests ($\tau$-Learner) and multitask GPs (Multitask-Learners) turn out to be quite reliable methods across different simulations, as they are the best performing or never lag far behind. However, we highlight that the appropriate choice of a Meta-Learner should be primarily based on domain knowledge about the specific study, since there is no effective and widely accepted empirical way to perform Meta-Learner model selection on CATE. The main drivers should include knowledge about strength of confounding and complexity of heterogeneity patterns. As discussed above in relation to simulated studies results, if the suspected complexity of CATE differs greatly from that of the prognostic score, $\mu (\bm{x}_i) = \mathbb{E} \big[ Y_i \mid \bm{X}_i = \bm{x}_i, Z_i = 0 \big]$, complexity, then X-, R-, and $\tau$-Learners are proved to be a more appropriate choice. In particular, R- and $\tau$-Learners allow to convey targeted regularization on CATE (in the form of shrinkage priors in the case of BCF) thanks to their different outcome model's parametrization. They also efficiently leverage propensity score estimates to deal with strong confounding patterns. In addition, $\tau$-Learners such as BCF, have the inherent advantage of returning full predictive posterior estimates, that can be handily exploited to measure uncertainty and lack of common support.

	\section{The effect of school meal programs on health indicators} \label{sec:realdata}
	\DIFaddend 
	
	In this section we provide a full-length analysis of the NHANES data introduced in Section \ref{sec:intro}, previously analyzed by \cite{chan_2016}, to demonstrate the use of CATE estimation methods and related tools in the study of heterogeneity. The dataset consists of $N=2330$ observations and $P=11$ covariates. The outcome variable of interest $Y_i$ is child's BMI, while the treatment $Z_i$ denotes participation in the National School Lunch or the School Breakfast programs, which are both designed to tackle poor or insufficient food access in low-income households. The full list of the variables, including the available covariates, is provided in Table \ref{NHANES} in the Appendix. This specific setup suggests that the impact of participation in school meal programs might be heterogeneous across children, in that demographics such as age, gender or ethnicity might play a role in how effective participation is (e.g.~younger kids might benefit more than older ones, etc.). This advocates the use of methods for CATE estimation.

	By taking a rather agnostic approach to the problem, we decide to employ Bayesian Causal Forests ($\tau$-Learner) \citep{hahn_2020} in the analysis, for two primary reasons. Together with the causal multitask GPs of \citep{vanderschaar_2017, vanderschaar_2018}, BCF was the most flexible method across different CATE simulations in the earlier section. In addition, as an advantage over causal multitask GPs, BCF yields direct posterior distribution estimates on CATE that can be easily used for quantifying uncertainty around point estimates, especially to tackle out-of-overlap predictions. The causal multitask GPs can also output posterior distribution estimates on the conditional average potential outcomes $\mu_Z(\bm{x}_i) = \mathbb{E} \big[ Y_i \mid Z_i = z_i, \bm{X}_i = \bm{x}_i \big]$, but as they are structured as T-Learners, and thus fit two models with independent error terms across $Z_i = z_i$, they do not output a natural measure of uncertainty around CATE directly (see \cite{hahn_2020} for a more detailed discussion of the issue).

	\DIFaddend \begin{figure}[t]
	\centering
	\begin{minipage}{.5\textwidth}
		\centering
		\includegraphics[width=1\linewidth]{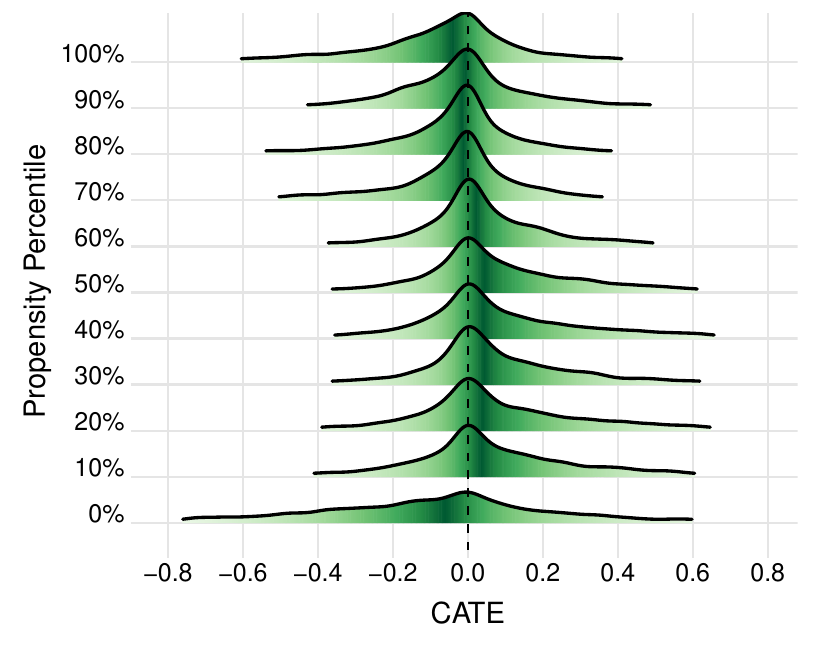}
	\end{minipage}%
	\begin{minipage}{.5\textwidth}
		\centering
		\includegraphics[width=1\linewidth]{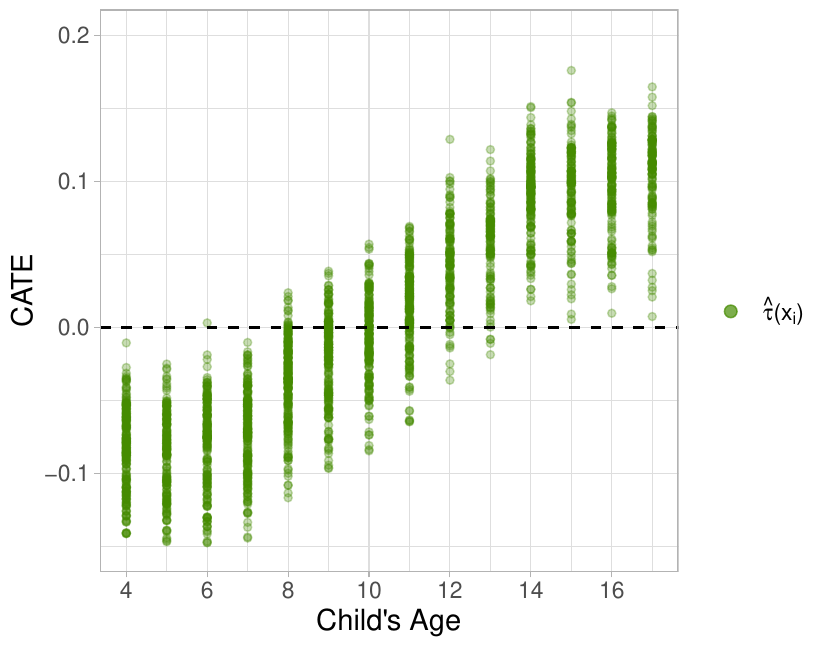}
	\end{minipage}
	\caption{\small Left pane: BCF's posterior distribution estimates on CATE corresponding to the approximated propensity score percentiles (i.e.~to individuals in the sample whose estimated propensity corresponds or is closest to PS percentiles) . Right pane: BCF's CATE point estimates (averaged over the $5\,000$ post burn-in MCMC iterations) as a function of child's age.}
	\label{figNHANES}
\end{figure}

	First of all, we employ a 1-hidden-layer neural network classifier to estimate the propensity score as a function of all the covariates, to be added as an additional covariate, as exclusively envisaged by the $\tau$-Learner framework. We then run Bayesian Causal Forest algorithm for a total of $10\,000$ MCMC iterations, of which the first $5\,000$ are discarded as burn-in. The left pane in Figure \ref{figNHANES} represents the resulting CATE posterior distributions corresponding to different approximated propensity score percentiles (namely to individuals in the sample whose estimated propensity is equal or closest to the PS percentiles). This type of plot allows us to visualize uncertainty in the form of how concentrated (or diffuse) are individual CATE estimates around their mean or median. The figure depicts CATE point estimates which are nearly zero for all propensity score levels, essentially corresponding to a null average treatment effect and very weak or absent heterogeneity patterns. Uncertainty quantification is more or less the same across PS percentiles, apart from the minimum (i.e.~0\% percentile) where CATE distribution is much more diffuse, potentially indicating inference in a poor overlap region in that instance. 
	
\begin{figure}
        \centering
        \includegraphics[scale=0.63]{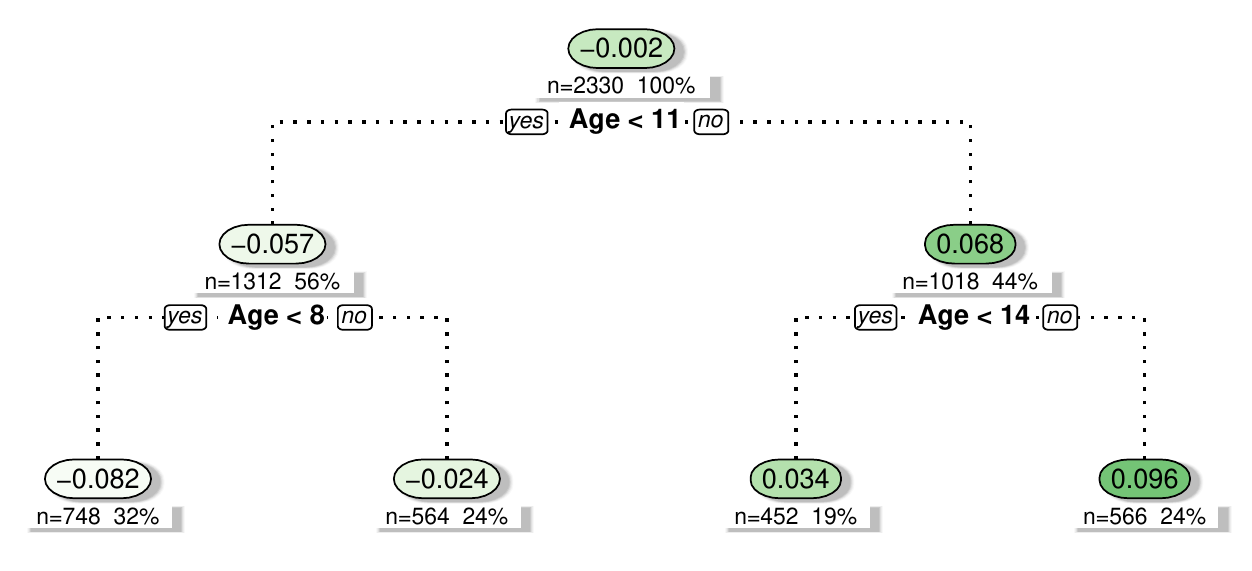}
        \caption{\small Decision tree indicating the most homogeneous subgroups in terms of treatment response, as a function of the available covariates (moderators). The nodes report CATE estimates averaged within the corresponding subgroup. The first node intuitively reports ATE estimate.}
        \label{fig:dectree}
    \end{figure}

	Drawing attention to the study of moderation effects, we run a simple decision tree partition algorithm using the \textsc{R} package \texttt{rpart}, where average CATE point estimates are treated as the target variable, while the covariates $\bm{X}_i \in \mathcal{X}$ are treated as predictors. The purpose of this exercise, which can be carried out using CATE estimates obtained from any Meta-Learner, is to find the most homogeneous subgroups in terms of treatment response and the most informative moderating covariates. The results from this exercise are depicted in Figure \ref{fig:dectree} in the form of a simple decision tree, where nodes report CATE estimates averaged within the corresponding subgroup, and provide evidence of very little, if not null, heterogeneity arising from Children's Age, given that the first two most informative splits in the tree feature this covariate, and that the estimated treatment effect is not very different across these subgroups. To better visualize this relationship, in the right pane of Figure \ref{figNHANES} we plot point estimates of CATE against Children's Age, that show a weak but positive relationship. Figures \ref{figNHANES} and \ref{fig:dectree} capture the role of Children's Age. Although it does not appear to be a major driver of propensity score (Ethnicity, Poverty Level and Participation to other Food Programs seem to be the main determinants of $Z$ --- see Table \ref{tab:logit} in the Appendix), it is likely the main source of the, albeit small, moderation effects.

	From the original analysis carried out by \cite{chan_2016} on the same NHANES dataset, it emerges that the estimated average treatment effect (ATE), on the 2007-2008 logged data, is significantly small, perhaps actually null. As stated by the authors, this finding is likely attributable to the fact that the school meal programs are already well implemented, that is, treatment administration is already targeting the right population of individuals, with the policy implication that there is no need for re-designing it. To relate their findings to our analysis, we notice that results are very similar, in that we find no significant treatment effect across propensity score percentiles (Figure \ref{figNHANES}), and neither across subgroups defined by children's age (Figure \ref{fig:dectree}). Treatment response patterns emerging from this analysis can be linked back to a setting similar to the ACTG simulated example, where CATE is weakly heterogeneous and of simple form (in this case virtually constant and null), such that S-, X- and $\tau$-Learners would be the preferred choice. Results from BCF, implementation of $\tau$-Learners, in this analysis demonstrate its particular feature of being able, as described by \cite{hahn_2020}, to shrink CATE estimates back to homogeneity if required, through targeted regularization.

	\section{Conclusion} \label{sec:discussion}
    
    In the previous sections, we discussed the most recent developments on estimation of heterogeneous treatment effects in the context of non-randomized studies. We use here the phrase ``non-randomized" studies as the methods illustrated are well suited for observational studies, but also for randomized experiments that might display threats to the randomization mechanism, such as imperfect compliance or non-random missingness.
	
	Our review of Meta-Learner frameworks and simulation studies lead to a few general observations. With regards to the properties of Meta-Learners reviewed in Section \ref{sec:metalearners}, there is a clear distinction between different groups of Meta-Learners according to the type of CATE complexity encountered. S-Learners are appropriate when CATE has simple heterogeneity patterns, while T-Learners when CATE is rather complex. The other Meta-Learners are instead designed to ideally adjust to different CATE complexity setups, although some of them appear to do so better than others. Multitask-Learners feature parameters sharing in the estimation of conditional average POs across $Z_i = z_i$, but being an extension of T-Learners they essentially assume separate models, with independent error terms, for the conditional average POs. For this reason, although they generally perform drastically better than T-Learners, they appear to do worse in settings with weak heterogeneity. The remaining batch, formed by X-, R- and $\tau$-Learners, consist of methods that are instead specifically designed to capture simple heterogeneity patterns, in different ways. X-Learners successfully extend T-Learners by applying propensity score re-weighting, while R- and $\tau$-Learners are both propensity methods that apply targeted regularization/shrinkage by exploiting \cite{robinson_1988} parametrization. 
	
	As for the results from the simulation exercises in Section \ref{sec:simu}, we described how BCF ($\tau$-Learners) and causal multitask GPs (Multitask-Learners) both performed consistently well across different simulated scenarios, with X-Learner trailing just a bit behind, and thus appear to be quite flexible and reliable methods to be chosen. BCF has also been shown to be particularly effective in addressing strong confounding. Finally, we stress that a practical advantage of R- and $\tau$-Learners over T-Learners and extensions (X- and Multitask-Learners) is that they are designed to tackle strong confounding and they can produce, in addition to point estimates, direct measures of uncertainty on CATE, which is equally essential in applied studies. A Bayesian implementation of R- and $\tau$-Learners also has the advantage of conveniently returning full posterior distributions on CATE, as demonstrated in Section \ref{sec:realdata} in the context of BCF.

    We conclude thus by recommending that the selection of a Meta-Learner should be primarily based on domain knowledge about the potential complexity of treatment effects heterogeneity patterns. If complex patterns are suspected, then a Multitask-Learner such as CMGP or NSGP is an appropriate choice. If simple-to-moderate CATE complexity, which we believe is more common in empirical studies, and strong confounding are suspected instead, X-, R- and $\tau$-Learners are more adequate, with the additional point in favor of R- and $\tau$-Learners that they provide more natural measure of uncertainty around CATE estimates.

	\nocite{*}
	\bibliographystyle{chicago}
	\bibliography{ReviewReferences}

	\clearpage

	\appendix
	
	\section{Supplementary Results on simulated examples}
	
	This appendix section contains additional results on the simulated exercises encountered in Section \ref{sec:IHDP} and \ref{sec:ACTG} respectively. In particular we report tables with $\mbox{Bias}_{\tau}$ estimates on CATT (in-ovelap regions), and $\mbox{Bias}_{\tau}$ and $\sqrt{\mbox{PEHE}}_{\tau}$ estimates on CATC (out-of-overlap regions).
	
		\begin{table}[!h]
		\caption{\small IHDP and ACTG-175 simulated exercises of Section \ref{sec:simu}. $\bm{\mbox{Bias}_{\tau}}$ estimates $\pm$ 95\% confidence intervals for each tested model on in-overlap \textbf{CATT}, on train and test sets respectively.}
		\label{tab:bias1}
		\footnotesize
		\centering
		\begin{tabular}{ccc c cc}
			\toprule
			&   \multicolumn{2}{c}{\textbf{IHDP}} & & \multicolumn{2}{c}{\textbf{ACTG-175}}  \\
			
			\cmidrule(r){2-3}  \cmidrule(r){5-6}
			
			& Train  &  Test &   & Train  &  Test  \\
			
			\midrule
			
			S-RF  		&	-0.05 $\pm$ 0.04		&  -0.07  $\pm$ 0.05  &   &  0.01 $\pm$ 0.01	&   0.00 $\pm$ 0.01  \\
			S-BART  		&	-0.10 $\pm$ 0.03 	&  -0.16 $\pm$ 0.04      & 	&  -0.03 $\pm$ 0.01	&  -0.03 $\pm$ 0.01  \\

			\midrule
			
			T-RF  		&	-0.02 $\pm$ 0.03		&  -0.01  $\pm$ 0.03  &   &  0.00 $\pm$ 0.01	&   -0.01 $\pm$ 0.01  \\
			T-BART  		&	-0.04 $\pm$ 0.02 	&  -0.03 $\pm$ 0.03      & 	&  0.01 $\pm$ 0.01	&  0.00 $\pm$ 0.01  \\
			
			\midrule
			
			X-RF  		&	-0.08 $\pm$ 0.03		&  -0.10  $\pm$ 0.05  &   &  -0.04 $\pm$ 0.01	&   -0.04 $\pm$ 0.01  \\
			X-BART  		&	0.06 $\pm$ 0.02 	&  0.04 $\pm$ 0.02      & 	&  -0.06 $\pm$ 0.01	 &  -0.06 $\pm$ 0.01  \\
			
			\midrule

			R-LASSO  		&	0.18 $\pm$ 0.03 	&  0.14 $\pm$ 0.03      &    &  0.01 $\pm$ 0.01	&   0.01 $\pm$ 0.01	\\
			R-BOOST  		&	0.22 $\pm$ 0.02 	&  0.23 $\pm$ 0.04      &    &  0.10 $\pm$ 0.01	&   0.12 $\pm$ 0.01	\\
			CF  		&	0.06 $\pm$ 0.03 	&  0.05 $\pm$ 0.05      &    &  -0.04 $\pm$ 0.01	&   -0.04 $\pm$ 0.01	\\
            
            \midrule
            
            CMGP  		&   \textbf{0.00 $\pm$ 0.01} 	&	\textbf{0.00 $\pm$ 0.02}      &	   &  -0.04 $\pm$ 0.01 	&	-0.05 $\pm$ 0.01	\\
			NSGP		&  	\textbf{0.00 $\pm$ 0.01} 	&	\textbf{0.00 $\pm$ 0.02}      &  	&  -0.04 $\pm$ 0.01 	&	-0.05 $\pm$ 0.01	\\
			
			\midrule
			
			BCF  		&	0.04 $\pm$ 0.02 	&  0.03 $\pm$ 0.03      &    &  \textbf{0.00 $\pm$ 0.01}	&   \textbf{0.00 $\pm$ 0.01}	\\
			
			\bottomrule
		\end{tabular}  
	\end{table} 	

		\begin{table}[!h]
		\caption{\small IHDP and ACTG-175 simulated exercises of Section \ref{sec:simu}. $\bm{\mbox{Bias}_{\tau}}$ estimates $\pm$ 95\% confidence intervals for each tested model, on out-of-overlap \textbf{CATC}, on train and test sets respectively.}
		\label{tab:bias2}
		\footnotesize
		\centering
		\begin{tabular}{ccc c cc}
			\toprule
			&   \multicolumn{2}{c}{\textbf{IHDP}} & & \multicolumn{2}{c}{\textbf{ACTG-175}}  \\
			
			\cmidrule(r){2-3}  \cmidrule(r){5-6}
			
			& Train  &  Test &   & Train  &  Test  \\
			
			\midrule
			
			S-RF  		&	-0.05 $\pm$ 0.05		&  -0.05  $\pm$ 0.05  &   &  -0.09 $\pm$ 0.01	&   -0.08 $\pm$ 0.01  \\
			S-BART  		&	-0.14 $\pm$ 0.03 	&  -0.13 $\pm$ 0.03      & 	&  -0.04 $\pm$ 0.01	&  -0.04 $\pm$ 0.01  \\

			\midrule
			
			T-RF  		&	-0.03 $\pm$ 0.01		&  -0.04  $\pm$ 0.02  &   &  -0.09 $\pm$ 0.01	&   -0.08 $\pm$ 0.01  \\
			T-BART  		&	\textbf{0.01 $\pm$ 0.01} 	&  \textbf{0.01 $\pm$ 0.01}      & 	&  -0.07 $\pm$ 0.01	&  -0.07 $\pm$ 0.01  \\
			
			\midrule
			
			X-RF  		&	0.14 $\pm$ 0.03		&  0.14  $\pm$ 0.04  &   &  -0.04 $\pm$ 0.01	&   -0.04 $\pm$ 0.01  \\
			X-BART  		&	-0.13 $\pm$ 0.03 	&  -0.13 $\pm$ 0.03      & 	&  -0.02 $\pm$ 0.01	 &  -0.02 $\pm$ 0.01  \\
			
			\midrule

			R-LASSO  		&	0.06 $\pm$ 0.02 	&  0.06 $\pm$ 0.02      &    &  0.02 $\pm$ 0.01	&   0.02 $\pm$ 0.01	\\
			R-BOOST  		&	0.41 $\pm$ 0.03 	&  0.40 $\pm$ 0.04      &    &  0.18 $\pm$ 0.01	&   0.16 $\pm$ 0.01	\\
			CF  		&	0.24 $\pm$ 0.04 	&  0.25 $\pm$ 0.04      &    &  \textbf{-0.01 $\pm$ 0.01}	&   \textbf{-0.01 $\pm$ 0.01}	\\
            
            \midrule
            
            CMGP  		&   0.11 $\pm$ 0.07 	&	0.10 $\pm$ 0.07     &	   &  -0.04 $\pm$ 0.01 	&	-0.04 $\pm$ 0.01	\\
			NSGP		&  	0.11 $\pm$ 0.02 	&	0.10 $\pm$ 0.03      &  	&  -0.05 $\pm$ 0.01 	&	-0.05 $\pm$ 0.01	\\
			
			\midrule
			
			BCF  		&	0.03 $\pm$ 0.01 	&  0.04 $\pm$ 0.02      &    &  \textbf{0.01 $\pm$ 0.01}	&   \textbf{0.01 $\pm$ 0.01}	\\
			
			\bottomrule
		\end{tabular}  
	\end{table} 		
	
	\clearpage
	
		\begin{table}[!h]
		\caption{\small IHDP and ACTG-175 simulated exercises of Section \ref{sec:simu}. $\bm{\sqrt{\mbox{\textbf{PEHE}}}_{\tau}}$ estimates $\pm$ 95\% confidence intervals for each tested model, on out-of-overlap \textbf{CATC}, on train and test sets respectively.}
		\label{tab:PEHEcatc}
		\footnotesize
		\centering
		\begin{tabular}{ccc c cc}
			\toprule
			&   \multicolumn{2}{c}{\textbf{IHDP}} & & \multicolumn{2}{c}{\textbf{ACTG-175}}  \\
			
			\cmidrule(r){2-3}  \cmidrule(r){5-6}
			
			& Train  &  Test &   & Train  &  Test  \\
			
			\midrule
			
			S-RF  		&	2.95 $\pm$ 0.25		&  3.16  $\pm$ 0.25  &   &  0.59 $\pm$ 0.01	&   0.51 $\pm$ 0.01  \\
			S-BART  		&	2.13 $\pm$ 0.14 	&  2.20 $\pm$ 0.14      & 	&  0.45 $\pm$ 0.01	&  0.46 $\pm$ 0.01  \\

			\midrule
			
			T-RF  		&	1.70 $\pm$ 0.12		&  2.39  $\pm$ 0.17  &   &  0.60 $\pm$ 0.01	&   0.51 $\pm$ 0.01  \\
			T-BART  		&	0.82 $\pm$ 0.02 	&  1.43 $\pm$ 0.09      & 	&  0.55 $\pm$ 0.01	&  0.55 $\pm$ 0.01  \\
			
			\midrule
			
			X-RF  		&	3.35 $\pm$ 0.23		&  3.39  $\pm$ 0.23  &   &  \textbf{0.36 $\pm$ 0.01}	&   \textbf{0.36 $\pm$ 0.01}  \\
			X-BART  		&	2.21 $\pm$ 0.15 	&  2.25 $\pm$ 0.15      & 	&  0.44 $\pm$ 0.01	 &  0.44 $\pm$ 0.01  \\
			
			\midrule

			R-LASSO  		&	2.02 $\pm$ 0.14 	&  2.07 $\pm$ 0.15      &    &  0.63 $\pm$ 0.01	&   0.63 $\pm$ 0.01	\\
			R-BOOST  		&	2.34 $\pm$ 0.15 	&  2.52 $\pm$ 0.16      &    &  0.52 $\pm$ 0.01	&   0.51 $\pm$ 0.01	\\
			CF  		&	3.14 $\pm$ 0.21 	&  3.07 $\pm$ 0.20      &    &  0.40 $\pm$ 0.01	&   0.40 $\pm$ 0.01	\\
            
            \midrule
            
            CMGP  		&   0.83 $\pm$ 0.09 	&	1.10 $\pm$ 0.11      &	   &  0.45 $\pm$ 0.01 	&	0.45 $\pm$ 0.01	\\
			NSGP		&  	\textbf{0.66 $\pm$ 0.03} 	&	\textbf{0.97 $\pm$ 0.09}      &  	&  0.43 $\pm$ 0.01 	&	0.43 $\pm$ 0.01	\\
			
			\midrule
			
			BCF  		&	0.73 $\pm$ 0.02 	&  1.35 $\pm$ 0.10      &    &  0.39 $\pm$ 0.01	&   0.39 $\pm$ 0.01	\\
			
			\bottomrule
		\end{tabular}  
	\end{table}

	\section{ACTG-175 data: a third simulated exercise} \label{sec:appendix}
	
	In this second short appendix section we present results obtained from a third semi-simulated exercise involving the ACTG-175 dataset. The structure of the utilized ACTG-175 data is exactly the same as the one found in the example in Section \ref{sec:ACTG} (same number of covariates, sample size, etc.). The only difference lies in how the prognostic score and CATE functions are simulated. For this third setup we chose slightly more complex surfaces compared to the ones in the other ACTG-175 simulation, to provide an additional example on the performance of the reviewed methods under a more challenging data generating process (closer to the one seen in the IHDP data example). More specifically, the two $\mu(\bm{x}_i)$ and $\tau(\bm{x}_i)$ surfaces are generated as
	\begin{equation}  \label{setup2}
	\begin{split}
	\mu(\bm{x}_i) = & ~ 6 + 0.3 wtkg^2 - \sin (age) \cdot (gender + 1) + 0.6 hemo \cdot race - 0.2z30 ~ ,
	\\[5pt]
	\tau(\bm{x}_i) = & ~ 1 + 1.5 \sin (wtkg) \cdot (karnof_{hi} + 1) + 0.4age^2 ~ .
	\end{split}
	\end{equation}
	Surfaces in (\ref{setup2}) feature more complex functions and more interaction terms. As in the other ACTG-175 data setup, Gaussian noise was added by simulating $\varepsilon_i \sim \mathcal{N} (0, \sigma^2)$, with standard deviation equal to $\sigma =   0.2 (\mu_{max} - \mu_{min})$, where $\mu_{max}$ is the sample maximum of the generated prognostic score, while $\mu_{min}$ is the sample minimum value.

	\begin{table}[!h]
		\caption{Third simulated setup (ACTG-175 data). $\bm{\sqrt{\mbox{\textbf{PEHE}}_{\tau}}}$ estimates $\pm$ 95\% confidence intervals for each tested model on in-overlap \textbf{CATT}, on the train and test sets respectively.}
		\label{aidsresults2}
		\footnotesize
		
		\centering
		
		\begin{tabular}{ccc}
			\toprule
			&   \multicolumn{2}{c}{\textbf{ACTG-175: 3$^{rd}$ simulation}} \\
			
			\cmidrule(r){2-3} 
			
			& Train  &  Test \\
			
			\midrule
			
			S-RF  		&	0.97 $\pm$ 0.01		&  1.03  $\pm$ 0.01     \\
			S-BART  		&	0.91 $\pm$ 0.01 	&  0.95 $\pm$ 0.01   \\

			\midrule
			
			T-RF  		&	0.82 $\pm$ 0.01		&  0.88  $\pm$ 0.01   \\
			T-BART  		&	0.84 $\pm$ 0.01 	&  0.93 $\pm$ 0.01    \\
			
			\midrule
			
			X-RF  		&	0.76 $\pm$ 0.01		&  0.81  $\pm$ 0.01   \\
			X-BART  		&	0.81 $\pm$ 0.01 	&  0.88 $\pm$ 0.01  \\
			
			\midrule

			R-LASSO  		&	1.13 $\pm$ 0.01 	&  1.18 $\pm$ 0.01  \\
			R-BOOST  		&	0.87 $\pm$ 0.01 	&  0.92 $\pm$ 0.01  \\
			CF  		&	0.87 $\pm$ 0.01 	&  0.87 $\pm$ 0.01 	\\
            
            \midrule
            
            CMGP  		&   0.61 $\pm$ 0.01 	&	0.72 $\pm$ 0.01   \\
			NSGP		&  	\textbf{0.59 $\pm$ 0.01} 	&	\textbf{0.70 $\pm$ 0.01}	\\
			
			\midrule
			
			BCF  		&	0.77 $\pm$ 0.01 	&  0.87 $\pm$ 0.01 	\\
			
			\bottomrule
		\end{tabular}
	\end{table}
	
	Results in terms of performance of the tested models are reported in Table \ref{aidsresults2}. A ranking similar to the one encountered in the IHDP data example emerges, with the GP Multitask-Learners being the best methods, followed up by X-RF and BCF. This is explained by the fact that CATE here is the result of a complex function, which tends to favour methods that fit separate surfaces (T-Learners, Multitask-Learners, etc.), just like in the IHDP example. Notice in fact that also in this case T-Learners show better performance than their S-Learner counterparts (T-RF vs S-RF, T-BART vs S-BART).

	\section{NHANES variables list}
	
	Table \ref{NHANES} contains the full list of variables included in the NHANES dataset analyzed in Section \ref{sec:realdata}.
	
	\begin{table}[!h]
	\caption{\small NHANES variables}
	\centering \footnotesize
	\begin{tabularx}{9.7cm}{l | X}
		
		\textbf{Variable}  &  \textbf{Description} \\
		\midrule
		
		\textit{BMI} &  Numeric. Body mass index (outcome variable)	\\	
		\textit{school\textunderscore meal} &	Binary (treatment indicator)	\\	
		\textit{age} &	 Numeric (child's age)	\\	
		\textit{childSex} &	 Binary (male = 1)	\\
		\textit{afam} & Binary (African American = 1)	\\
		\textit{hisam} & Binary (Hispanic = 1) \\
		\textit{povlev\textunderscore 200} & Binary (family above 200\% federal poverty lvl = 1)	\\
		\textit{sup\textunderscore nutr} &	Binary (supplementary nutrition program = 1)	\\	
		\textit{stamp\textunderscore prog} & Binary (food stamp program = 1)	\\	
		\textit{food\textunderscore sec} &	Binary (food security in household = 1)	\\
		\textit{ins} &	 Binary (any insurance = 1)   \\
		\textit{refsex} &	Binary (adult respondent gender is male = 1)  \\
		\textit{refage} &	Numeric (adult respondent's age)	
		
		\label{NHANES}
	\end{tabularx}
\end{table}

\begin{table}[!htbp] \centering \small
  \caption{Logit regression model of $Z$ as a function of the covariates $\bm{X}$. Coefficients display log odds ratio. Stars indicate level of significance. Ethnicity (African America, Hispanic), Poverty Level and participation to other food programs (Food Stamp) appear to have the greatest and most significant impact on selection into treatment. Child's Age (the main moderator) is significant but of smaller magnitude.} 
  \label{tab:logit} 
\begin{tabular}{@{\extracolsep{5pt}}lc} 
\\[-1.8ex]\hline 
\hline \\[-1.8ex] 
 & \multicolumn{1}{c}{\textit{Dependent variable:}} \\ 
\cline{2-2} 
\\[-1.8ex] & Z \\ 
\hline \\[-1.8ex] 
 Child's Age & 0.052$^{***}$ \\ 
  & (0.013) \\[7pt]
 Ref Age & 0.001 \\ 
  & (0.005) \\[7pt]
 Child's Sex & $-$0.010 \\ 
  & (0.098) \\[7pt]
 African & 1.047$^{***}$ \\ 
  & (0.123) \\[7pt]
 Hispanic & 1.086$^{***}$ \\ 
  & (0.123) \\[7pt]
 Poverty Lvl & $-$1.407$^{***}$ \\ 
  & (0.110) \\[7pt]
 Suppl Nutr & 0.244$^{*}$ \\ 
  & (0.140) \\[7pt]
 Food Stamp & 1.117$^{***}$ \\ 
  & (0.131) \\[7pt]
 Food Security & 0.345$^{***}$ \\ 
  & (0.122) \\[7pt]
 Insurance & $-$0.021 \\ 
  & (0.143) \\[7pt]
 Ref Sex & 0.023 \\ 
  & (0.102) \\[7pt]
 Constant & $-$0.669$^{**}$ \\ 
  & (0.275) \\[7pt]
\hline \\[-1.8ex] 
Observations & 2,330 \\ 
Log Likelihood & $-$1,260.824 \\ 
Akaike Inf. Crit. & 2,545.647 \\ 
\hline 
\hline \\[-1.8ex] 
\textit{Note:}  & \multicolumn{1}{r}{$^{*}$p$<$0.1; $^{**}$p$<$0.05; $^{***}$p$<$0.01} \\ 
\end{tabular} 
\end{table}

\end{document}